%%%%%%%%%%%%%% TEX FILE OF AN ACCEPTED PAPER AT Int. J. Mod. Phys. E %%%
%%%%%%%%%%%%%% ACCEPTED MANUSCRIPT NUMBER 9207.004 %%%%%%%%%%%%%%%%%%%%%
%%%%%%%%%%%%%% ACCEPTING EDITOR: BRUCE McKELLAR %%%%%%%%%%%%%%%%%%%%%%%%%
%%%%%%%%%%%%%%%%%%%%%%%%%%%%%%%%%%%%%%%%%%%%%%%%%%%%%%%%%%%%%%%%%%%%%%%%%
% % % % % % % % % % % % % % % % % % % % % % % % % % % % % % % % % % % %
%%%   This is PHYZZX macro package.   % % % % % % % % % % % % % % % % %
%% % % % % % % % % % % % % % % % % % % % % % % % % % % % % % % % % % % %
%%%  This version of PHYZZX should be used with Version 1.0 of TEX  % %
%% % % % % % % % % % % % % % % % % % % % % % % % % % % % % % % % % % % %
%%%   Do not "\input phyzzx" unless you preload or "\input" PLAIN.  % %
%% % % % % % % % % % % % % % % % % % % % % % % % % % % % % % % % % % % %
%%%   To preload both PLAIN and PHYZZX, begin your file with    % % % %
%%%  a line "%macropackage=phyzzx" instead of "\input phyzzx".  % % % %
%% % % % % % % % % % % % % % % % % % % % % % % % % % % % % % % % % % % %
%%%%%%%%%%%%%%%%%%%%%%%%%%%%%%%%%%%%%%%%%%%%%%%%%%%%%%%%%%%%%%%%%%%%%%%%
%%%%%%%  Created by Vadim Kaplunovsky in June 1984.   %%%%%%%%%%%%%%%%%%
% % % % % % % % % % % % % % % % % % % % % % % % % % % % % % % % % % % %
%%%%%%%%%%%%  Latest update/debug: May 29, 1985   %%%%%%%%%%%%%%%%%%%%%%
%%%%%%%%%%%%%%%%%%%%%%%%%%%%%%%%%%%%%%%%%%%%%%%%%%%%%%%%%%%%%%%%%%%%%%%%
%
%graphics
%
\def\gronk#1#2#3
    {\midinsert
         \noindent\hfil\hbox to #1truein
              {\vbox to #2truein
                   {\special{include(#3 origin nocarriage)}\vfill}
              \hfil}
    \hfil\endinsert}
%
% am fonts changed to cm fonts and syntax of special command changed
% to conform with new release of TeX from Texas A&M - E. Groth  12-Sep-86
%
\expandafter\ifx\csname phyzzx\endcsname\relax\else
 \errhelp{Hit <CR> and go ahead.}
 \errmessage{PHYZZX macros are already loaded or input. }
 \endinput \fi
\catcode`\@=11 % This allows us to modify PLAIN macros.
%
%%%%%%%%%%%%%%%%%%%%%%%%%%%%%%%%%%%%%%%%%%%%%%%%%%%%%%%%%%%%%%%%%%%%%%%%
%
%   I begin with fonts.
%
\font\seventeenrm=cmr10 scaled\magstep3
\font\fourteenrm=cmr10 scaled\magstep2
\font\twelverm=cmr10 scaled\magstep1
\font\ninerm=cmr9            \font\sixrm=cmr6

\font\fourteenbf=cmbx10 scaled\magstep2
\font\twelvebf=cmbx10 scaled\magstep1
\font\ninebf=cmbx9            \font\sixbf=cmbx6
\font\seventeeni=cmmi10 scaled\magstep3     \skewchar\seventeeni='177
\font\fourteeni=cmmi10 scaled\magstep2      \skewchar\fourteeni='177
\font\twelvei=cmmi10 scaled\magstep1        \skewchar\twelvei='177
\font\ninei=cmmi9                           \skewchar\ninei='177
\font\sixi=cmmi6                            \skewchar\sixi='177
\font\seventeensy=cmsy10 scaled\magstep3    \skewchar\seventeensy='60
\font\fourteensy=cmsy10 scaled\magstep2     \skewchar\fourteensy='60
\font\twelvesy=cmsy10 scaled\magstep1       \skewchar\twelvesy='60
\font\ninesy=cmsy9                          \skewchar\ninesy='60
\font\sixsy=cmsy6                           \skewchar\sixsy='60

\font\fourteenex=cmex10 scaled\magstep2
\font\twelveex=cmex10 scaled\magstep1
%\font\elevenex=cmex10 scaled\magstephalf
%

\font\fourteensl=cmsl10 scaled\magstep2
\font\twelvesl=cmsl10 scaled\magstep1
\font\ninesl=cmsl9

\font\fourteenit=cmti10 scaled\magstep2
\font\twelveit=cmti10 scaled\magstep1
\font\nineit=cmti9
\font\fourteentt=cmtt10 scaled\magstep2
\font\twelvett=cmtt10 scaled\magstep1
\font\fourteencp=cmcsc10 scaled\magstep2
\font\twelvecp=cmcsc10 scaled\magstep1
\font\tencp=cmcsc10
\newfam\cpfam
\newdimen\b@gheight        \b@gheight=12pt
\newcount\f@ntkey        \f@ntkey=0
\def\f@m{\afterassignment\samef@nt\f@ntkey=}
\def\samef@nt{\fam=\f@ntkey \the\textfont\f@ntkey\relax}
\def\rm{\f@m0 }
\def\mit{\f@m1 }         
\def\cal{\f@m2 }
\def\it{\f@m\itfam}
\def\sl{\f@m\slfam}
\def\bf{\f@m\bffam}
\def\tt{\f@m\ttfam}
\def\caps{\f@m\cpfam}
\def\fourteenpoint{\relax
    \textfont0=\fourteenrm          \scriptfont0=\tenrm
      \scriptscriptfont0=\sevenrm
    \textfont1=\fourteeni           \scriptfont1=\teni
      \scriptscriptfont1=\seveni
    \textfont2=\fourteensy          \scriptfont2=\tensy
      \scriptscriptfont2=\sevensy
    \textfont3=\fourteenex          \scriptfont3=\twelveex
      \scriptscriptfont3=\tenex
    \textfont\itfam=\fourteenit     \scriptfont\itfam=\tenit
    \textfont\slfam=\fourteensl     \scriptfont\slfam=\tensl
    \textfont\bffam=\fourteenbf     \scriptfont\bffam=\tenbf
      \scriptscriptfont\bffam=\sevenbf
    \textfont\ttfam=\fourteentt
    \textfont\cpfam=\fourteencp
    \samef@nt
    \b@gheight=14pt
    \setbox\strutbox=\hbox{\vrule height 0.85\b@gheight
                depth 0.35\b@gheight width\z@ }}
\def\twelvepoint{\relax
    \textfont0=\twelverm          \scriptfont0=\ninerm
      \scriptscriptfont0=\sixrm
    \textfont1=\twelvei           \scriptfont1=\ninei
      \scriptscriptfont1=\sixi
    \textfont2=\twelvesy           \scriptfont2=\ninesy
      \scriptscriptfont2=\sixsy
    \textfont3=\twelveex          \scriptfont3=\tenex
      \scriptscriptfont3=\tenex
    \textfont\itfam=\twelveit     \scriptfont\itfam=\nineit
    \textfont\slfam=\twelvesl     \scriptfont\slfam=\ninesl
    \textfont\bffam=\twelvebf     \scriptfont\bffam=\ninebf
      \scriptscriptfont\bffam=\sixbf
    \textfont\ttfam=\twelvett
    \textfont\cpfam=\twelvecp
    \samef@nt
    \b@gheight=12pt
    \setbox\strutbox=\hbox{\vrule height 0.85\b@gheight
                depth 0.35\b@gheight width\z@ }}
\def\tenpoint{\relax
    \textfont0=\tenrm          \scriptfont0=\sevenrm
      \scriptscriptfont0=\fiverm
    \textfont1=\teni           \scriptfont1=\seveni
      \scriptscriptfont1=\fivei
    \textfont2=\tensy          \scriptfont2=\sevensy
      \scriptscriptfont2=\fivesy
    \textfont3=\tenex          \scriptfont3=\tenex
      \scriptscriptfont3=\tenex
    \textfont\itfam=\tenit     \scriptfont\itfam=\seveni
    \textfont\slfam=\tensl     \scriptfont\slfam=\sevenrm
    \textfont\bffam=\tenbf     \scriptfont\bffam=\sevenbf
      \scriptscriptfont\bffam=\fivebf
    \textfont\ttfam=\tentt
    \textfont\cpfam=\tencp
    \samef@nt
    \b@gheight=10pt
    \setbox\strutbox=\hbox{\vrule height 0.85\b@gheight
                depth 0.35\b@gheight width\z@ }}
%
%%%%%%%%%%%%%%%%%%%%%%%%%%%%%%%%%%%%%%%%%%%%%%%%%%%%%%%%%%%%%%%%%%%%%%%%
%
%   Next, I define basic spacing parameters.
%
\normalbaselineskip = 20pt plus 0.2pt minus 0.1pt
\normallineskip = 1.5pt plus 0.1pt minus 0.1pt
\normallineskiplimit = 1.5pt
\newskip\normaldisplayskip
\normaldisplayskip = 20pt plus 5pt minus 10pt
\newskip\normaldispshortskip
\normaldispshortskip = 6pt plus 5pt
\newskip\normalparskip
\normalparskip = 6pt plus 2pt minus 1pt
\newskip\skipregister
\skipregister = 5pt plus 2pt minus 1.5pt
\newif\ifsingl@    \newif\ifdoubl@
\newif\iftwelv@    \twelv@true
\def\singlespace{\singl@true\doubl@false\spaces@t}
\def\doublespace{\singl@false\doubl@true\spaces@t}
\def\normalspace{\singl@false\doubl@false\spaces@t}
\def\Tenpoint{\tenpoint\twelv@false\spaces@t}
\def\Twelvepoint{\twelvepoint\twelv@true\spaces@t}
\def\spaces@t{\relax
      \iftwelv@ \ifsingl@\subspaces@t3:4;\else\subspaces@t1:1;\fi
       \else \ifsingl@\subspaces@t3:5;\else\subspaces@t4:5;\fi \fi
      \ifdoubl@ \multiply\baselineskip by 5
         \divide\baselineskip by 4 \fi }
\def\subspaces@t#1:#2;{
      \baselineskip = \normalbaselineskip
      \multiply\baselineskip by #1 \divide\baselineskip by #2
      \lineskip = \normallineskip
      \multiply\lineskip by #1 \divide\lineskip by #2
      \lineskiplimit = \normallineskiplimit
      \multiply\lineskiplimit by #1 \divide\lineskiplimit by #2
      \parskip = \normalparskip
      \multiply\parskip by #1 \divide\parskip by #2
      \abovedisplayskip = \normaldisplayskip
      \multiply\abovedisplayskip by #1 \divide\abovedisplayskip by #2
      \belowdisplayskip = \abovedisplayskip
      \abovedisplayshortskip = \normaldispshortskip
      \multiply\abovedisplayshortskip by #1
        \divide\abovedisplayshortskip by #2
      \belowdisplayshortskip = \abovedisplayshortskip
      \advance\belowdisplayshortskip by \belowdisplayskip
      \divide\belowdisplayshortskip by 2
      \smallskipamount = \skipregister
      \multiply\smallskipamount by #1 \divide\smallskipamount by #2
      \medskipamount = \smallskipamount \multiply\medskipamount by 2
      \bigskipamount = \smallskipamount \multiply\bigskipamount by 4 }
\def\normalbaselines{ \baselineskip=\normalbaselineskip
   \lineskip=\normallineskip \lineskiplimit=\normallineskip
   \iftwelv@\else \multiply\baselineskip by 4 \divide\baselineskip by 5
     \multiply\lineskiplimit by 4 \divide\lineskiplimit by 5
     \multiply\lineskip by 4 \divide\lineskip by 5 \fi }

\Twelvepoint  % That's the default
\interlinepenalty=50
\interfootnotelinepenalty=5000
\predisplaypenalty=9000
\postdisplaypenalty=500
\hfuzz=1pt
\vfuzz=0.2pt
\voffset=0pt
\dimen\footins=8 truein
%
%%%%%%%%%%%%%%%%%%%%%%%%%%%%%%%%%%%%%%%%%%%%%%%%%%%%%%%%%%%%%%%%%%%%%%%%
%
%   Next, I define output routines, footnotes & related stuff.
%
\def\pagecontents{
   \ifvoid\topins\else\unvbox\topins\vskip\skip\topins\fi
   \dimen@ = \dp255 \unvbox255
   \ifvoid\footins\else\vskip\skip\footins\footrule\unvbox\footins\fi
   \ifr@ggedbottom \kern-\dimen@ \vfil \fi }
\def\makeheadline{\vbox to 0pt{ \skip@=\topskip
      \advance\skip@ by -12pt \advance\skip@ by -2\normalbaselineskip
      \vskip\skip@ \line{\vbox to 12pt{}\the\headline} \vss
      }\nointerlineskip}
\def\makefootline{\baselineskip = 1.5\normalbaselineskip
                 \line{\the\footline}}
\newif\iffrontpage
\newif\ifletterstyle
\newif\ifp@genum
\def\nopagenumbers{\p@genumfalse}
\def\pagenumbers{\p@genumtrue}
\pagenumbers
\newtoks\paperheadline
\newtoks\letterheadline
\newtoks\paperfootline
\newtoks\letterfootline
\newtoks\letterinfo
\newtoks\Letterinfo
\newtoks\date
\footline={\ifletterstyle\the\letterfootline\else\the\paperfootline\fi}
\paperfootline={\hss\iffrontpage\else\ifp@genum\tenrm\folio\hss\fi\fi}
\letterfootline={\iffrontpage\the\letterinfo\else\hfil\fi}
\Letterinfo={\hfil}
\letterinfo={\hfil}
\headline={\ifletterstyle\the\letterheadline\else\the\paperheadline\fi}
\paperheadline={\hfil}
\letterheadline{\iffrontpage \hfil \else
    \rm \ifp@genum page \ \folio\fi \hfil\the\date \fi}
\def\monthname{\relax\ifcase\month 0/\or January\or February\or
   March\or April\or May\or June\or July\or August\or September\or
   October\or November\or December\else\number\month/\fi}
\def\today{\monthname\ \number\day, \number\year}
\date={\today}
\countdef\pageno=1      \countdef\pagen@=0
\countdef\pagenumber=1  \pagenumber=1
\def\advancepageno{\global\advance\pagen@ by 1
   \ifnum\pagenumber<0 \global\advance\pagenumber by -1
    \else\global\advance\pagenumber by 1 \fi \global\frontpagefalse }
\def\folio{\ifnum\pagenumber<0 \romannumeral-\pagenumber
           \else \number\pagenumber \fi }
\def\footrule{\dimen@=\prevdepth\nointerlineskip
   \vbox to 0pt{\vskip -0.25\baselineskip \hrule width 0.35\hsize \vss}
   \prevdepth=\dimen@ }
\newtoks\foottokens
\foottokens={}
\newdimen\footindent
\footindent=24pt
\def\vfootnote#1{\insert\footins\bgroup
   \interlinepenalty=\interfootnotelinepenalty \floatingpenalty=20000
   \singl@true\doubl@false\Tenpoint
   \splittopskip=\ht\strutbox \boxmaxdepth=\dp\strutbox
   \leftskip=\footindent \rightskip=\z@skip
   \parindent=0.5\footindent \parfillskip=0pt plus 1fil
   \spaceskip=\z@skip \xspaceskip=\z@skip
   \the\foottokens
   \Textindent{$ #1 $}\footstrut\futurelet\next\fo@t}
\def\Textindent#1{\noindent\llap{#1\enspace}\ignorespaces}
\def\footnote#1{\attach{#1}\vfootnote{#1}}

\let\footsymbol=\star
\newcount\lastf@@t           \lastf@@t=-1
\newcount\footsymbolcount    \footsymbolcount=0
\newif\ifPhysRev
\def\bumpfootsymbolcount{\relax
   \iffrontpage \bumpfootsymbolNP \else \advance\lastf@@t by 1
     \ifPhysRev \bumpfootsymbolPR \else \bumpfootsymbolNP \fi \fi
   \global\lastf@@t=\pagen@ }
\def\bumpfootsymbolNP{\ifnum\footsymbolcount <0 \global\footsymbolcount =0 \fi
    \ifnum\lastf@@t<\pagen@ \global\footsymbolcount=0
     \else \global\advance\footsymbolcount by 1 \fi }
\def\bumpfootsymbolPR{\ifnum\footsymbolcount >0 \global\footsymbolcount =0 \fi
      \global\advance\footsymbolcount by -1 }
\def\fd@f#1 {\xdef\footsymbol{\mathchar"#1 }}
\def\generatefootsymbol{\ifcase\footsymbolcount \fd@f 13F \or \fd@f 279
    \or \fd@f 27A \or \fd@f 278 \or \fd@f 27B \else
    \ifnum\footsymbolcount <0 \fd@f{023 \number-\footsymbolcount }
     \else \fd@f 203 {\loop \ifnum\footsymbolcount >5
        \fd@f{203 \footsymbol } \advance\footsymbolcount by -1
        \repeat }\fi \fi }

\def\nonfrenchspacing{\sfcode`\.=3001 \sfcode`\!=3000 \sfcode`\?=3000
    \sfcode`\:=2000 \sfcode`\;=1500 \sfcode`\,=1251 }
\nonfrenchspacing
\newdimen\d@twidth
{\setbox0=\hbox{s.} \global\d@twidth=\wd0 \setbox0=\hbox{s}
    \global\advance\d@twidth by -\wd0 }
\def\removehglue{\loop \unskip \ifdim\lastskip >\z@ \repeat }
\def\roll@ver#1{\removehglue \nobreak \count255 =\spacefactor \dimen@=\z@
    \ifnum\count255 =3001 \dimen@=\d@twidth \fi
    \ifnum\count255 =1251 \dimen@=\d@twidth \fi
    \iftwelv@ \kern-\dimen@ \else \kern-0.83\dimen@ \fi
   #1\spacefactor=\count255 }
\def\step@ver#1{\relax \ifmmode #1\else \ifhmode
    \roll@ver{${}#1$}\else {\setbox0=\hbox{${}#1$}}\fi\fi }
\def\attach#1{\step@ver{\strut^{\mkern 2mu #1} }}
%
%%%%%%%%%%%%%%%%%%%%%%%%%%%%%%%%%%%%%%%%%%%%%%%%%%%%%%%%%%%%%%%%%%%%%%%%
%
%   Here come chapter, section, subsection & appendix macros.
%
\newcount\chapternumber      \chapternumber=0
\newcount\sectionnumber      \sectionnumber=0
\newcount\equanumber         \equanumber=0
\let\chapterlabel=\relax
\let\sectionlabel=\relax
\newtoks\chapterstyle        \chapterstyle={\Number}
\newtoks\sectionstyle        \sectionstyle={\chapterlabel\Number}
\newskip\chapterskip         \chapterskip=\bigskipamount
\newskip\sectionskip         \sectionskip=\medskipamount
\newskip\headskip            \headskip=8pt plus 3pt minus 3pt
\newdimen\chapterminspace    \chapterminspace=15pc
\newdimen\sectionminspace    \sectionminspace=10pc
\newdimen\referenceminspace  \referenceminspace=25pc
\def\chapterreset{\global\advance\chapternumber by 1
   \ifnum\equanumber<0 \else\global\equanumber=0\fi
   \sectionnumber=0 \makechapterlabel}
\def\makechapterlabel{\let\sectionlabel=\relax
   \xdef\chapterlabel{\the\chapterstyle{\the\chapternumber}.}}
\def\alphabetic#1{\count255='140 \advance\count255 by #1\char\count255}
\def\Alphabetic#1{\count255='100 \advance\count255 by #1\char\count255}
\def\Roman#1{\uppercase\expandafter{\romannumeral #1}}
\def\roman#1{\romannumeral #1}
\def\Number#1{\number #1}
\def\BLANC#1{}
\def\titlestyle#1{\par\begingroup \interlinepenalty=9999
     \leftskip=0.02\hsize plus 0.23\hsize minus 0.02\hsize
     \rightskip=\leftskip \parfillskip=0pt
     \hyphenpenalty=9000 \exhyphenpenalty=9000
     \tolerance=9999 \pretolerance=9000
     \spaceskip=0.333em \xspaceskip=0.5em
     \iftwelv@\fourteenpoint\else\twelvepoint\fi
   \noindent #1\par\endgroup }
\def\spacecheck#1{\dimen@=\pagegoal\advance\dimen@ by -\pagetotal
   \ifdim\dimen@<#1 \ifdim\dimen@>0pt \vfil\break \fi\fi}
\def\TableOfContentEntry#1#2#3{\relax}
\def\chapter#1{\par \penalty-300 \vskip\chapterskip
   \spacecheck\chapterminspace
   \chapterreset \titlestyle{\chapterlabel\ #1}
   \TableOfContentEntry c\chapterlabel{#1}
   \nobreak\vskip\headskip \penalty 30000
   \wlog{\string\chapter\space \chapterlabel} }

\def\section#1{\par \ifnum\the\lastpenalty=30000\else
   \penalty-200\vskip\sectionskip \spacecheck\sectionminspace\fi
   \global\advance\sectionnumber by 1
   \xdef\sectionlabel{\the\sectionstyle\the\sectionnumber}
   \wlog{\string\section\space \sectionlabel}
   \TableOfContentEntry s\sectionlabel{#1}
   \noindent {\caps\enspace\sectionlabel\quad #1}\par
   \nobreak\vskip\headskip \penalty 30000 }
\def\subsection#1{\par
   \ifnum\the\lastpenalty=30000\else \penalty-100\smallskip \fi
   \noindent\undertext{#1}\enspace \vadjust{\penalty5000}}

\def\undertext#1{\vtop{\hbox{#1}\kern 1pt \hrule}}
\def\APPENDIX#1#2{\par\penalty-300\vskip\chapterskip
   \spacecheck\chapterminspace \chapterreset \xdef\chapterlabel{#1}
   \titlestyle{APPENDIX #2} \nobreak\vskip\headskip \penalty 30000
   \TableOfContentEntry a{#1}{#2}
   \wlog{\string\Appendix\ \chapterlabel} }
\def\Appendix#1{\APPENDIX{#1}{#1}}
\def\appendix{\APPENDIX{A}{}}
\def\unnumberedchapters{\let\makechapterlabel=\relax \let\chapterlabel=\relax
   \sectionstyle={\BLANC}\let\sectionlabel=\relax \sequentialequations }
%
%%%%%%%%%%%%%%%%%%%%%%%%%%%%%%%%%%%%%%%%%%%%%%%%%%%%%%%%%%%%%%%%%%%%%%%%
%
%   Here come macros for equation numbering.
%
\def\eqname#1{\relax \ifnum\equanumber<0
     \xdef#1{{\noexpand\rm(\number-\equanumber)}}%
       \global\advance\equanumber by -1
    \else \global\advance\equanumber by 1
      \xdef#1{{\noexpand\rm(\chapterlabel\number\equanumber)}} \fi #1}

\def\eqn{\eqno\eqname}

\def\eqinsert#1{\noalign{\dimen@=\prevdepth \nointerlineskip
   \setbox0=\hbox to\displaywidth{\hfil #1}
   \vbox to 0pt{\kern 0.5\baselineskip\hbox{$\!\box0\!$}\vss}
   \prevdepth=\dimen@}}
%

%
%%%%%%%%%%%%%%%%%%%%%%%%%%%%%%%%%%%%%%%%%%%%%%%%%%%%%%%%%%%%%%%%%%%%%%%%
%   Here come items and lists
%
\def\GENITEM#1;#2{\par \hangafter=0 \hangindent=#1
    \Textindent{$ #2 $}\ignorespaces}
\outer\def\newitem#1=#2;{\gdef#1{\GENITEM #2;}}
\newdimen\itemsize                \itemsize=30pt
\newitem\item=1\itemsize;
\newitem\sitem=1.75\itemsize;     
\newitem\ssitem=2.5\itemsize;     
\outer\def\newlist#1=#2&#3&#4;{\toks0={#2}\toks1={#3}%
   \count255=\escapechar \escapechar=-1
   \alloc@0\list\countdef\insc@unt\listcount     \listcount=0
   \edef#1{\par
      \countdef\listcount=\the\allocationnumber
      \advance\listcount by 1
      \hangafter=0 \hangindent=#4
      \Textindent{\the\toks0{\listcount}\the\toks1}}
   \expandafter\expandafter\expandafter
    \edef\c@t#1{begin}{\par
      \countdef\listcount=\the\allocationnumber \listcount=1
      \hangafter=0 \hangindent=#4
      \Textindent{\the\toks0{\listcount}\the\toks1}}
   \expandafter\expandafter\expandafter
    \edef\c@t#1{con}{\par \hangafter=0 \hangindent=#4 \noindent}
   \escapechar=\count255}
\def\c@t#1#2{\csname\string#1#2\endcsname}
\newlist\point=\Number&.&1.0\itemsize;
\newlist\subpoint=(\alphabetic&)&1.75\itemsize;
\newlist\subsubpoint=(\roman&)&2.5\itemsize;
%

%
%%%%%%%%%%%%%%%%%%%%%%%%%%%%%%%%%%%%%%%%%%%%%%%%%%%%%%%%%%%%%%%%%%%%%%%%
%
%   Here come macros for references, figures & tables.
%
% % % % % % % % % % % % % % % % % % % % % % % % % % % % % % % % % % % %
%%  First, references.
%
\newcount\referencecount     \referencecount=0
\newcount\lastrefsbegincount \lastrefsbegincount=0
\newif\ifreferenceopen       \newwrite\referencewrite
\newif\ifrw@trailer
\newdimen\refindent     \refindent=30pt
\def\NPrefmark#1{\attach{\scriptscriptstyle [ #1 ] }}
\let\PRrefmark=\attach
\def\refmark#1{\relax\ifPhysRev\PRrefmark{#1}\else\NPrefmark{#1}\fi}
\def\refend@{\refmark{\number\referencecount}}
\def\refend{\refend@{}\space }
\def\refsend{\refmark{\count255=\referencecount
   \advance\count255 by-\lastrefsbegincount
   \ifcase\count255 \number\referencecount
   \or \number\lastrefsbegincount,\number\referencecount
   \else \number\lastrefsbegincount-\number\referencecount \fi}\space }
\def\refitem#1{\par \hangafter=0 \hangindent=\refindent \Textindent{#1}}
\def\Ref{\rw@trailertrue\REF}
\def\ref{\Ref\?}

\def\REF#1{\r@fstart{#1}%
   \rw@begin{\the\referencecount.}\rw@end}
\def\REFS#1{\r@fstart{#1}%
   \lastrefsbegincount=\referencecount
   \rw@begin{\the\referencecount.}\rw@end}
\def\r@fstart#1{\chardef\rw@write=\referencewrite \let\rw@ending=\refend@
   \ifreferenceopen \else \global\referenceopentrue
   \immediate\openout\referencewrite=referenc.txa
   \toks0={\catcode`\^^M=10}\immediate\write\rw@write{\the\toks0} \fi
   \global\advance\referencecount by 1 \xdef#1{\the\referencecount}}
{\catcode`\^^M=\active %
 \gdef\rw@begin#1{\immediate\write\rw@write{\noexpand\refitem{#1}}%
   \begingroup \catcode`\^^M=\active \let^^M=\relax}%
 \gdef\rw@end#1{\rw@@end #1^^M\rw@terminate \endgroup%
   \ifrw@trailer\rw@ending\global\rw@trailerfalse\fi }%
 \gdef\rw@@end#1^^M{\toks0={#1}\immediate\write\rw@write{\the\toks0}%
   \futurelet\n@xt\rw@test}%
 \gdef\rw@test{\ifx\n@xt\rw@terminate \let\n@xt=\relax%
       \else \let\n@xt=\rw@@end \fi \n@xt}%
}
\let\rw@ending=\relax
\let\rw@terminate=\relax
\let\splitout=\relax
\def\Closeout#1{\toks0={\catcode`\^^M=5}\immediate\write#1{\the\toks0}%
   \immediate\closeout#1}
%
% % % % % % % % % % % % % % % % % % % % % % % % % % % % % % % % % % % %
%%  Next, figure captions and table captions.
%
\newcount\figurecount     \figurecount=0
\newcount\tablecount      \tablecount=0
\newif\iffigureopen       \newwrite\figurewrite
\newif\iftableopen        \newwrite\tablewrite
\def\FIG#1{\f@gstart{#1}%
   \rw@begin{\the\figurecount)}\rw@end}

\def\Fig{\rw@trailertrue\def\rw@ending{Fig.~\?}\FIG\?}
\def\fig{\rw@trailertrue\def\rw@ending{fig.~\?}\FIG\?}
\def\TABLE#1{\T@Bstart{#1}%
   \rw@begin{\the\tableecount:}\rw@end}
\def\Table{\rw@trailertrue\def\rw@ending{Table~\?}\TABLE\?}
\def\f@gstart#1{\chardef\rw@write=\figurewrite
   \iffigureopen \else \global\figureopentrue
   \immediate\openout\figurewrite=figures.txa
   \toks0={\catcode`\^^M=10}\immediate\write\rw@write{\the\toks0} \fi
   \global\advance\figurecount by 1 \xdef#1{\the\figurecount}}
\def\T@Bstart#1{\chardef\rw@write=\tablewrite
   \iftableopen \else \global\tableopentrue
   \immediate\openout\tablewrite=tables.txa
   \toks0={\catcode`\^^M=10}\immediate\write\rw@write{\the\toks0} \fi
   \global\advance\tablecount by 1 \xdef#1{\the\tablecount}}
%

%
% % % % % % % % % % % % % % % % % % % % % % % % % % % % % % % % % % % %
%%  Finally, inserted figures.
%
%\newread\figureread                                     %% That's
%\def\g@tfigure#1#2 {\openin\figureread #2.fig           %% an example
%   \ifeof\figureread \errhelp=\disabledfigures          %% of \g@tfigure
%     \errmessage{No such file: #2.fig}\let#1=\relax \else
%    \read\figureread to\y@p \read\figureread to\y@p     %%
%    \read\figureread to\x@p \read\figureread to\y@m     %% See LOCPHYX.TEX
%    \read\figureread to\x@m \closein\figureread         %% file for the
%    \xdef#1{\hbox{\kern-\x@m truein \vbox{\kern-\y@m truein
%      \hbox to \x@p truein{\vbox to \y@p truein{        %% actual definition.
%        \special{include(#2.fig relative)}\vss }\hss }}}}\fi }    %%
%
\def\getfigure#1{\global\advance\figurecount by 1
   \xdef#1{\the\figurecount}\count255=\escapechar \escapechar=-1
   \edef\n@xt{\noexpand\g@tfigure\csname\string#1Body\endcsname}%
   \escapechar=\count255 \n@xt }
\def\g@tfigure#1#2 {\errhelp=\disabledfigures \let#1=\relax
   \errmessage{\string\getfigure\space disabled}}
\newhelp\disabledfigures{ Empty figure of zero size assumed.}
\def\figinsert#1{\midinsert\Tenpoint\medskip
   \count255=\escapechar \escapechar=-1
   \edef\n@xt{\csname\string#1Body\endcsname}
   \escapechar=\count255 \centerline{\n@xt}
   \bigskip\narrower\narrower
   \noindent{\it Figure}~#1.\quad }
%
%%%%%%%%%%%%%%%%%%%%%%%%%%%%%%%%%%%%%%%%%%%%%%%%%%%%%%%%%%%%%%%%%%%%%%%%
%
%   Here come macros for memos & letters.
%
\def\masterreset{\global\pagenumber=1 \global\chapternumber=0
   \global\equanumber=0 \global\sectionnumber=0
   \global\referencecount=0 \global\figurecount=0 \global\tablecount=0 }
\def\FRONTPAGE{\ifvoid255\else\vfill\penalty-20000\fi
      \masterreset\global\frontpagetrue
      \global\lastf@@t=0 \global\footsymbolcount=0}

\def\paperstyle{\letterstylefalse\normalspace\papersize}
\def\letterstyle{\letterstyletrue\singlespace\lettersize}
\def\papersize{\hsize=35 truepc\vsize=50 truepc\hoffset=2 truepc
               \skip\footins=\bigskipamount}
\def\lettersize{\hsize=6.5 truein\vsize=8.5 truein\hoffset=0 truein
   \skip\footins=\smallskipamount \multiply\skip\footins by 3 }
\paperstyle   %  This is the default
%
% % % % % % % % % % % % % % % % % % % % % % % % % % % % % % % % % % % %
%
\def\MEMO{\letterstyle \letterinfo={\hfil } \let\rule=\memorule
    \FRONTPAGE \memohead }
\let\memohead=\relax

\def\memit@m#1{\smallskip \hangafter=0 \hangindent=1in
      \Textindent{\caps #1}}
\def\subject{\memit@m{Subject:}}
\def\topic{\memit@m{Topic:}}
\def\from{\memit@m{From:}}
\def\to{\relax \ifmmode \rightarrow \else \memit@m{To:}\fi }
\def\memorule{\medskip\hrule height 1pt\bigskip}
\newwrite\labelswrite
\newtoks\rw@toks

\def\addressee#1{\medskip\rightline{\the\date\hskip 30pt} \bigskip
   \ialign to\hsize{\strut ##\hfil\tabskip 0pt plus \hsize \cr #1\crcr}
   \writelabel{#1}\medskip\par\noindent}
\def\rwl@begin#1\cr{\rw@toks={#1\crcr}\relax
   \immediate\write\labelswrite{\the\rw@toks}\futurelet\n@xt\rwl@next}
\def\rwl@next{\ifx\n@xt\rwl@end \let\n@xt=\relax
      \else \let\n@xt=\rwl@begin \fi \n@xt}
\let\rwl@end=\relax
\def\writelabel#1{\immediate\write\labelswrite{\noexpand\labelbegin}
     \rwl@begin #1\cr\rwl@end
     \immediate\write\labelswrite{\noexpand\labelend}}
\newbox\FromLabelBox

\let\labelend=\relax
\def\labelbegin#1\labelend{\setbox0=\vbox{\ialign{##\hfil\cr #1\crcr}}
     \MakeALabel }
\newtoks\FromAddress
\FromAddress={}
\def\MakeFromBox#1{\global\setbox\FromLabelBox=\vbox{\Tenpoint
     \ialign{##\hfil\cr #1\the\FromAddress\crcr}}}
\newdimen\labelwidth        \labelwidth=6in
\def\MakeALabel{\vskip 1pt \hbox{\vrule \vbox{
    \hsize=\labelwidth \hrule\bigskip
    \leftline{\hskip 1\parindent \copy\FromLabelBox}\bigskip
    \centerline{\hfil \box0 } \bigskip \hrule
    }\vrule } \vskip 1pt plus 1fil }
\newskip\signatureskip       \signatureskip=30pt
\def\signed#1{\par \penalty 9000 \medskip \dt@pfalse
  \everycr={\noalign{\ifdt@p\vskip\signatureskip\global\dt@pfalse\fi}}
  \setbox0=\vbox{\singlespace \ialign{\strut ##\hfil\crcr
   \noalign{\global\dt@ptrue}#1\crcr}}
  \line{\hskip 0.5\hsize minus 0.5\hsize \box0\hfil} \medskip }
\newbox\letterb@x
\def\lettertext{\par\unvcopy\letterb@x\par}
\def\multiletter{\setbox\letterb@x=\vbox\bgroup
      \everypar{\vrule height 1\baselineskip depth 0pt width 0pt }
      \singlespace \topskip=\baselineskip }
\def\letterend{\par\egroup}
%
%%%%%%%%%%%%%%%%%%%%%%%%%%%%%%%%%%%%%%%%%%%%%%%%%%%%%%%%%%%%%%%%%%%%%%%
%
%   Here come macros for title pages.
%
\newskip\frontpageskip
\newtoks\Pubnum
\newtoks\pubtype
\newif\ifp@bblock  \p@bblocktrue
\def\PH@SR@V{\doubl@true \baselineskip=24.1pt plus 0.2pt minus 0.1pt
             \parskip= 3pt plus 2pt minus 1pt }
\def\PHYSREV{\paperstyle\PhysRevtrue\PH@SR@V}
\def\titlepage{\FRONTPAGE\paperstyle\ifPhysRev\PH@SR@V\fi
   \ifp@bblock\p@bblock \else\hrule height\z@ \relax \fi }
\def\nopubblock{\p@bblockfalse}
\def\endpage{\vfil\break}
\frontpageskip=12pt plus .5fil minus 2pt
\pubtype={\tensl Preliminary Version}
\Pubnum={}
\def\p@bblock{\begingroup \tabskip=\hsize minus \hsize
   \baselineskip=1.5\ht\strutbox \topspace-2\baselineskip
   \halign to\hsize{\strut ##\hfil\tabskip=0pt\crcr
       \the\Pubnum\crcr\the\date\crcr\the\pubtype\crcr}\endgroup}
\def\title#1{\vskip\frontpageskip \titlestyle{#1} \vskip\headskip }
\def\author#1{\vskip\frontpageskip\titlestyle{\twelvecp #1}\nobreak}

\def\address#1{\par\kern 5pt\titlestyle{\twelvepoint\it #1}}
\def\andaddress{\par\kern 5pt \centerline{\sl and} \address}

\def\abstract{\par\dimen@=\prevdepth \hrule height\z@ \prevdepth=\dimen@
   \vskip\frontpageskip\centerline{\fourteenrm ABSTRACT}\vskip\headskip }

%
%
%%%%%%%%%%%%%%%%%%%%%%%%%%%%%%%%%%%%%%%%%%%%%%%%%%%%%%%%%%%%%%%%%%%%%%%%
%   Miscellaneous macros
%

\def\\{\relax \ifmmode \backslash \else {\tt\char`\\}\fi }
\def\sequentialequations{\relax\if\equanumber<0\else\global\equanumber=-1\fi}

\def\journal#1&#2(#3){\unskip, \sl #1\unskip~\bf\ignorespaces #2\rm (19#3),}

\def\topspace{\hrule height 0pt depth 0pt \vskip}

\def\Buildrel#1\under#2{\mathrel{\mathop{#2}\limits_{#1}}}
\def\becomes#1{\mathchoice{\becomes@\scriptstyle{#1}}{\becomes@\scriptstyle
   {#1}}{\becomes@\scriptscriptstyle{#1}}{\becomes@\scriptscriptstyle{#1}}}
\def\becomes@#1#2{\mathrel{\setbox0=\hbox{$\m@th #1{\,#2\,}$}%
    \mathop{\hbox to \wd0 {\rightarrowfill}}\limits_{#2}}}

\let\int=\intop         
\def\lsim{\mathrel{\mathpalette\@versim<}}
\def\gsim{\mathrel{\mathpalette\@versim>}}
\def\@versim#1#2{\vcenter{\offinterlineskip
    \ialign{$\m@th#1\hfil##\hfil$\crcr#2\crcr\sim\crcr } }}
\def\big#1{{\hbox{$\left#1\vbox to 0.85\b@gheight{}\right.\n@space$}}}
\def\Big#1{{\hbox{$\left#1\vbox to 1.15\b@gheight{}\right.\n@space$}}}
\def\bigg#1{{\hbox{$\left#1\vbox to 1.45\b@gheight{}\right.\n@space$}}}
\def\Bigg#1{{\hbox{$\left#1\vbox to 1.75\b@gheight{}\right.\n@space$}}}
%
% % % % % % % % % % % % % % % % % % % % % % % % % % % % % % % % % % % %
%
%   Finally, some bug fixings.
%
\let\sec@nt=\sec
\def\sec{\relax\ifmmode\let\n@xt=\sec@nt\else\let\n@xt\section\fi\n@xt}
\def\obsolete#1{\message{Macro \string #1 is obsolete.}}
\def\firstsec#1{\obsolete\firstsec \section{#1}}
\def\firstsubsec#1{\obsolete\firstsubsec \subsection{#1}}
\def\thispage#1{\obsolete\thispage \global\pagenumber=#1\frontpagefalse}
\def\thischapter#1{\obsolete\thischapter \global\chapternumber=#1}
\def\REFSCON{\obsolete\REFSCON\REF}
\def\splitout{\obsolete\splitout\relax}
\def\prop{\obsolete\prop \propto }
\def\nextequation#1{\obsolete\nextequation \global\equanumber=#1
   \ifnum\the\equanumber>0 \global\advance\equanumber by 1 \fi}
\def\BOXITEM{\afterassigment\B@XITEM\setbox0=}
\def\B@XITEM{\par\hangindent\wd0 \noindent\box0 }
\def\phyzzx{PHY\setbox0=\hbox{Z}\copy0 \kern-0.5\wd0 \box0 X}
%
%%%%%%%%%%%%%%%%%%%%%%%%%%%%%%%%%%%%%%%%%%%%%%%%%%%%%%%%%%%%%%%%%%%%%%%%
%   That's about it
%
\everyjob{\xdef\today{\monthname\ \number\day, \number\year}}
\Pubnum={}
\pubtype={}
 at 7 truept
 at 10. truept
 at 12.00 truept

%
%\FromAddress={\crcr Joseph Henry Laboratories\cr Princeton University\cr
%    Princeton, New Jersey 08544\crcr}
\FromAddress={\crcr Center for Theoretical Physics\cr Texas A\&M University\cr
    College Station, Texas 77843-4242\crcr}
%
%\Letterinfo={\ninerm \hfil Telephone: (609) 452--4400\qquad
%                      Telex: 499--3512\hfil }
%\Letterinfo={\ninerm \hfil Bitnet: AHLUWALIA@TAMPHYS\qquad
%                      Fax: (409) 845--2590\hfil }
\Letterinfo={\ninerm \hfil  Fax: (505) 665--1712\hfil }

\edef\memorule{\medskip\hrule\kern 2pt\hrule \noindent
      \llap{\vbox to 0pt{ \vskip 1in\normalbaselines \tabskip=0pt plus 1fil
        \halign to 0.99in{\seventeenrm\hfil ##\hfil\cr
          M\cr E\cr M\cr O\cr R\cr A\cr N\cr D\cr U\cr M\cr}
        \vss }}\par\medskip}
\newread\figureread
\def\g@tfigure#1#2 {\openin\figureread #2.fig \ifeof\figureread
    \errmessage{No such file: #2.fig}\xdef#1{\hbox{}}\else
    \read\figureread to\y@p \read\figureread to\y@p
    \read\figureread to\x@p \read\figureread to\y@m
    \read\figureread to\x@m \closein\figureread
    \xdef#1{\hbox{\kern-\x@m truein \vbox{\kern-\y@m truein
      \hbox to \x@p truein{\vbox to \y@p truein{
        \special{include(#2.fig relative)}\vss }\hss }}}}\fi }
\catcode`\@=12 % at signs are no longer letters
%
%\message{ by V.K., G.S.}
        
%

\font\eightrm=cmr8
\font\eightbf=cmbx8

\font\eightsl=cmsl8
\font\eightmus=cmmi8
\def\smalltype{\let\rm=\eightrm \let\bf=\eightbf
\let\it=\eighti \let\sl=\eightsl \let\mus=\eightmus
\baselineskip=9.5pt minus .75pt
\rm}
\def\today{\ifcase\month\or
  January\or February\or March\or April\or May\or June\or
  July\or August\or September\or October\or November\or December\fi
  \space\number\day, \number\year}

\def\v#1{\vec#1}
\def\ra{\rangle}
\def\la{\langle}

\def\vt{\vert}

\def\pl{\partial}
\def\sqr#1#2{{\vcenter{\vbox{\hrule height.#2pt
        \hbox{\vrule width.#2pt height#1pt \kern#1pt
          \vrule width.#2pt}
        \hrule height.#2pt}}}}

\singlespace
%\doublespace
\def\mo{$m\rightarrow 0$~}

\def\v#1{\vec#1}
\def\ra{\rangle}
\def\la{\langle}

\def\vt{\vert}

\def\pl{\partial}
\def\sqr#1#2{{\vcenter{\vbox{\hrule height.#2pt
        \hbox{\vrule width.#2pt height#1pt \kern#1pt
          \vrule width.#2pt}
        \hrule height.#2pt}}}}

%%%%%%%%%%%%%%%%%%%%%%%%%%%%%%%  REFERENCES  %%%%%%%%%%%%%%%%%%%%%%%%%%%
\REF\CorbenS{H. C. Corben and J. Schwinger, { Phys. Rev.},
{\bf 58}, 953 (1940).}

\REF\JohnsonS{K. Johnson and E. C. G. Sudarshan, { Ann. Phys.} {\bf 13},
126 (1961).}

\REF\KobayashiS{M. Kobayashi and A. Shamaly, { Phys. Rev. D} {\bf 17},
2179 (1978).}

\REF\KobayashiTA{M. Kobayashi and Y. Takahashi, { Prog. Theor. Phys.} 
{\bf 75}, 993 (1986).}

\REF\KobayashiTB{M. Kobayashi and Y. Takahashi, { J. Phys. A} 
{\bf 20}, 6581 (1987).}

\REF\PRCb{D. V. Ahluwalia, D. J. Ernst and C. Burgard, 
submitted for publication.}

\REF\Thesis{D. V. Ahluwalia, Texas A\&M University Ph.~D. thesis (1991),
unpublished.}

\REF\Wigsym{D. V. Ahluwalia and D. J. Ernst, in
{\it ``Classical and Quantum Systems -- 
Foundations and Symmetries. 
Proceedings of the II International Wigner Symposium,''} Goslar, 1991, FRG 
(in press).}

\REF\Vanderbilt{D. V.  
Ahluwalia and D. J. Ernst,  in {\it ``Proceedings of the Computational
Quantum Physics Conference,''} Vanderbilt University, 1991 (in press).}

\REF\Pdata{J. J. Hernandez {\it et al.}, Phys. Lett B {\bf 239},  1 (1990).}

\REF\Gibbs{W. R. Gibbs {\it in} ``Proceedings of Physics with Light Mesons
and Second International Workshop on $\Pi$N Physics,'' (W. R. Gibbs
and B. M. K. Nefkins, Ed.) Los Alamos National Laboratory Report
LA--11184--C.}

\REF\Heavy{G. Bauer and C. A. Bertulani, { Phys. Rep.} {\bf 161} (1988)
299; J. S. Wu, C. Bottcher, M. R. Strayer and A. K. Kerman, { Ann.
Phys. (N.Y.)} {\bf 210} (1991) 402; M. Grabiak, B. M\"uller, W. Greiner, 
G. Soff and P. Koch, { J. Phys. G} {\bf15} (1989), L25.}

\REF\Charly{C. Bottcher, 
M. R. Strayer, C. J. Albert and D. J. Ernst, { Phys. Lett. B} {\bf 237}
(1990), 175.}

\REF\Jpsi{L. K\"opke and N. Wermes, { Phys. Rep.} {\bf 174} (1989),67.}

\REF\Fsup{``Glueballs, Hybrids and Exotic Hadrons, American Physics Institute
Conference Proceedings no.~185,'' (Suh--Urg Chung, Ed.) AIP, New York, 1989.}

\REF\DiracB {P. A. M. Dirac, { Proc. Roy. Soc. (London) A} {\bf 155}, 447
 (1936).}

\REF\DiracA{ P. A. M. Dirac,  Proc. Roy. Soc. (London) A {\bf 117},
610 (1928); {\it ibid.} {\bf 118},\rm 351 (1928).}

\REF\Anderson{ H. L. Anderson, E. Fermi, E. A. Long, R. Martin
and D. E. Nagle, { Phys. Rev. } {\bf 85}, 934 (1952);
E. Fermi, H. L. Anderson, A. Lundby, D. E. Nagle and G. B. Yodh,
{ Phys. Rev. } {\bf 85}, 935 (1952);
H. L. Anderson, E. Fermi, E. A. Long
and D. E. Nagle, { Phys. Rev. } {\bf 85}, 936 (1952).}

\REF\Wigner{E. P. Wigner, { Ann. of Math.} {\bf 40}, 149 (1939).}

\REF\WeinbergA{ S. Weinberg, { Phys. Rev. B} {\bf  133}, 1318 (1964).}

\REF\Joos{H. Joos, { Fortschr. Physik} {\bf 10}, 65 (1962).}

\REF\Duffin{ R. G. Duffin, { Phys. Rev.} {\bf 54}, 1114 (1938).}

\REF\Kemmer {N. Kemmer, { Proc. Roy. Soc. (London) A} {\bf 173}, 91 (1939).}

\REF\Fierz{ M. Fierz, { Helv. Phys. Acta} {\bf 12}, 3 (1939).}

\REF\FierzPA{ M. Fierz and W. Pauli, { Helv. Phys. Acta} {\bf 12}, 297
 (1939).}

\REF\FierzPB{ M. Fierz and W. Pauli, { Proc. Roy. Soc. (London) A} 
{\bf 173}, 211 (1939).} 

\REF\RaritaS{ W. Rarita and J. Schwinger, { Phys. Rev.}  {\bf 60}, \rm 
61 (1941).}

\REF\Bhabha{ H. J. Bhabha, { Rev. Mod. Phys.} {\bf 17}, 200 (1945).}

\REF\HarishA{Harish-Chandra, { Phys. Rev.} {\bf 71}, 793 (1947);
{ Proc. Roy. Soc. (London) A} {\bf 192}, 195
(1947).}

\REF\Bargmann{ V. Bargmann and E. P. Wigner, { Proc. Nat. Acad. Sci. (USA)}
{\bf 34}, 211 (1948).}

\REF\Foldy{L. L. Foldy, { Phys. Rev.} {\bf 102}, 568 (1956).}

\REF\SimilarA{D. L. Weaver, C. L. Hammer and R. H. Good, Jr., { Phys. Rev B}
{\bf 135}, 241 (1964).}

\REF\SimilarB{R. Shaw, { Nuovo Cimento A} {\bf 33}, 1074 (1964);
{\it ibid} {37}, 1086 (1965).}

\REF\SimilarC{D. L. Pursey, { Ann. Phys.} {\bf 32}, 157 (1965).}

\REF\SimilarD{A. McKerrel, { Proc. Roy. Soc. A} {\bf 285}, 287 (1965); 
{ Ann. Phys.} {\bf 40}, 237 (1966).}

\REF\Many{M. A. K. Khalil, { J. Phys. A} {\bf 12}, 649 (1979);
{ P. M. Mathews, B. Vijayalakshmi, M. Seetharaman, and
Y. Takahashi, { J. Phys. A} {\bf 14}, 1193 (1981)};
{ A. Z. Capri, {  Prog. Theor. Phys.} {\bf 48}, 1364 (1972);}
{ A. Amar and U. Dozzio, { Lett. Nuovo. Cim.} {\bf 5} 355 (1972)};
{J. Prabhakaran, T. R. Govindarajan, and M. Seetharaman,
{ Nucl. Phys. B} {\bf 127}, 537 (1977)};
{W. Becker, { J. Phys. A} {\bf 9}, 149 (1976)};
{ A. S. Glass, { Commun. Math. Phys.} {\bf 23}, 176 (1971).};
{ P. M. Mathews, M. Seetharaman, T. R. Govindarajan and
T. Prabhakaran, { Phys. Rev. D} {\bf 19}, 2947 (1979)}.}
\REF\KrajickN{R. A. Krajcik and M. M. Nieto, { Phys. Rev. D} {\bf 10}, 4049
(1974); {\it ibid.} {\bf 11}, 1442 (1975); {\it ibid.} {\bf 11}, 1459 (1975);
{\it ibid.} {\bf 13}, 924 (1976).} 

\REF\SimilarE{R. H. Good, Jr. { Ann. Phys.} {\bf 196}, 1 (1989).}

\REF\Ryder{L. H. Ryder, ``Quantum Field Theory,'' Cambridge University
Press, 1987.}

\REF\Mishra{V. K. Mishra, S. Hama, B. C. Clark, R. E. Kozack, R. L. Mercer
and L. Ray. { Phys. Rev. C} {\bf 43} (1991), 801; F. D. Santos, { Phys.
Lett. B} {\bf 175} (1986), 110; F. D. Santos and H. van Dam, { Phys. Rev. C}
{\bf 34} (1985), 250; H. M. Ruck and W. Griener,
{ J. Phys. G} {\bf 3} (1977), 657.}

%\REF\Poly{B. D. Keister and W. N. Polyzou, {\it in} ``Advances in Nuclear 
%Physics,'' (J. W. Negele and E. Vogt, Eds.) Plenum, New York, in press.}

\REF\Bjorken{J. D. Bjorken and S. D. Drell, 
``Relativistic Quantum Mechanics,'' Mc-Graw-Hill Book Co., New York, 1964.}

\REF\WeinbergF{S. Weinberg, `` Gravitation and Cosmology: Principles and
Applications of the General Theory of Relativity,'' Wiley \& Sons, 
New York, 1972.} 

\REF\WeinbergC{S.~Weinberg, {\it in }~~``Brandeis University Summer 
Institute in
Theoretical  Physics --- Lectures on Fields and Particles,''  (S. Deser and
K. W. Ford, Eds.), Vol. 2, p. 405,
Prentice Hall, Englewood Cliffs, N. J., 1965.} 

\REF\WeinbergQ{S. Weinberg, { Nucl. Phys. B (Proc. Suppl.)} {\bf 6}
(1989), 67.}

\REF\Ramond{P. Ramond, ``Field Theory: A Modern Primer,'' Addison-Wesley Publ.,
Redwood city, California, 1989.}

\REF\Schiff{L. I. Schiff, ``Quantum Mechanics,'' p. 203, McGraw--Hill Book
Co., New York,  1968.} 

\REF\WeinbergZ{S. Weinberg, { Phys. Rev. } {\bf  181}, 1893 (1969).}

\REF\Meson{M. B. Johnson and D. J. Ernst, Ann. Phys. 
(N.Y.) (in press).}

\REF\brief{D. V. Ahluwalia and D. J. Ernst, Phys. Rev. C {\bf 45}, 3010 (1992).}

\REF\Threehalf{D. V. Ahluwalia and D. J. Ernst, submitted for publication.}

\REF\WeinbergB{ S. Weinberg, { Phys. Rev. B} {\bf  134}, 882 (1964).}

\REF\Paradox{D. V. Ahluwalia and D. J. Ernst, Mod. Phys. Lett. A. 
{\bf 7}, 1967 (1992).}

\REF\Waveeqn{D. V. Ahluwalia and D. J. Ernst, Phys. Lett. B. {\bf 287}, 18, 
(1992).}
%%%%%%%%%%%%%%%%%%%%%%%%%%%%%%%%%%%%%%%%%%%%%%%%%%%%%%%%%%%%%%%%%%
\title{{$(j,0)\oplus(0,j)$ 
COVARIANT SPINORS AND CAUSAL PROPAGATORS
BASED ON WEINBERG FORMALISM}\footnote{*}{Work
supported, in part, by the U.S. National Science Foundation.}}
\rm 
%\vskip 0.5in
\author{ D.~V.~Ahluwalia\footnote{\dag}
{Present Address: Medium Energy Physics Division, MS H-850,
Los Alamos National Laboratory, Los Alamos, New Mexico 87545, U. S. A.}
and D. ~J.~Ernst\footnote{\ddag}{Present Address: 
Department of Physics and Astronomy
Vanderbilt University, Nashville, Tennessee 37235, U. S. A.}}

\address{Department of Physics and Center for Theoretical Physics,
Texas A\&M University, College Station, Texas 77843, U. S. A.}

\vfill\eject

\abstract
\noindent A pragmatic approach to constructing a covariant 
phenomenology of the interactions
of composite, high--spin hadrons is proposed. Because there are no known
wave equations without significant problems, we propose to construct
the phenomenology without explicit reference to a wave equation.
This is done by constructing the individual pieces of a perturbation theory
and then utilizing the perturbation theory as the definition of the 
phenomenology. The covariant spinors for a particle of spin $j$
are constructed directly from Lorentz invariance and the basic precepts
of quantum mechanics following the logic put forth originally by Wigner
and developed by Weinberg. Explicit expressions for the spinors
are derived for $j=~$1, 3/2 and 2. Field operators are constructed
from the spinors and the free--particle propagator is derived from
the vacuum expectation value of the time--order product of the field
operators. A few simple examples of model interactions are given.
This provides all the necessary ingredients to treat at a phenomenological
level and in a covariant manner particles of arbitrary spin.
\endpage

The study of quantum field theories of high--spin particles
has a long history. Despite considerable work and progress, there
remain fundamental difficulties [\CorbenS--\Vanderbilt] with each of
the various theoretical approaches which have so far been proposed. 
At the same time, there is a need
for an internally consistent way of treating high--spin objects in a
Lorentz covariant manner. For example, electroproduction of highly excited 
baryons in the nuclear medium can be studied at CEBAF. This work could be
extended at future accelerators such as KAON or PILAC
at LAMPF. The known baryons [\Pdata] have spin $j$ in the range 
$1/2\le j \le 15/2$. Higher spins might certainly be found in the future.
Studying
the final state interaction of an excited baryon with the residual
nucleus [\Gibbs], even at a qualitative level, could yield new insights 
into the quark structure of these excited hadrons. An alternate application
might be to use the strong electromagnetic
fields of the relativistic heavy ions at RHIC [\Heavy, \Charly]
to produce high--spin mesons (mesons with $j\le 6$ have been found)
through the two--photon mechanism.
Of particular interest might be the $f_2$(1720) (a meson with $j=2$)
whose structure might be predominantly glue as it is seen [\Jpsi]
in `gluon--rich' radiative decays of the $J/\Psi$ and it has a much
suppressed [\Fsup] electromagnetic coupling. 

The history of quantum field theories of high--spin particles is much
too extensive to review here. The first work on the subject
is that of Dirac [\DiracB], published eight years after his classic work 
[\DiracA]
on spin one--half particles. In this 1936 paper Dirac makes the following
observation, ``All the same, it is desirable to have the equations ready
for a possible future discovery of an elementary particle with spin
greater than a half, or for approximate application to composite particles.
Further, the underlying theory is of considerable mathematical interest.''
Sixteen years later, in a series of back--to--back
letters in {\it Physical Review}, Anderson and Fermi {\it et al.} [\Anderson]
reported the existence of an `intermediate state' of spin 3/2, the
$\Delta^{3\over 2} (1232)$. Since then, there have been many contributions
to the field. A representative sample of the work can be found in 
Refs.~[\Wigner--\SimilarE] and the history can be traced through the
references contained therein. 

Here, we make a pragmatic proposal for treating composite high--spin
particles in an internally consistent and Lorentz covariant manner.
We make use of the observation made by Wigner [\Wigner] and developed
extensively by Weinberg [\WeinbergA] that covariant spinors and field operators
follow directly from the basic precepts of quantum mechanics and
Poincar\'e covariance. In this work, we generalize the approach of Ryder
[\Ryder] to cast the work of
Weinberg[\WeinbergA] into a form which allows us to generate explicit
expressions for covariant spinors for particles of arbitrary spin. We here
produce explicit
expressions for spinors
with $j=1$, 3/2 and 2. For $j=1/2$ the procedure does of course,
reproduce the standard Dirac spinors. This demonstrates the practicality
of this approach and provides the needed spinors for our future
phenomenological work. The construction of the free--field operators from
the covariant spinors follows exactly the same logic as can be used
for the Dirac case. The free--particle propagator can be defined in terms
of the time--ordered product of the field operators. This is the
definition of the propagator which is required for a perturbation theory.
We here provide an explicit expression for the propagator so defined.
To this, we may add model interactions. These interactions we will take
from the simple Lorentz scalars that can be constructed from the
field operators and kinematical quantities available for the particular
problem being investigated. These interactions will include phenomenological
form factors in order to model the compositeness of the interacting
hadrons. Combining the covariant spinors, field operators and propagators
with the model interactions produces a well--defined perturbation theory.
We propose to use this perturbation theory as the basic definition
of our phenomenology. 

In this construction, we make no reference to any wave equation or to
any Lagrangian. This is less than an ideal circumstance. However,
there does not exist a wave equation for high--spin particles
[\CorbenS--\Vanderbilt] which does not have a fundamental difficulty.
Thus, in order to make progress at the phenomenological level, we
propose an end run around this difficulty --- working without a wave equation.
Clearly the lack of a wave
equation and a Lagrangian formulation might limit the applicability
and generality of our approach. 

One might ask, since we make extensive use of the work of Weinberg 
[\WeinbergA], why do we not utilize the Joos--Weinberg [\WeinbergA,\Joos]
equations? In investigating this possibility, we have found [\PRCb--\Vanderbilt]
that the Joos--Weinberg equations, even in the absence of
interactions, support unphysical solutions, a situation
which we term kinematic acausality. With this difficulty at the free--particle
level, attempts to introduce interactions into these equations [\Mishra]
can be  problematic because the interactions could mix in the unphysical
solutions.

The essential role of the Poincar\'e group in constructing Lorentz covariant
quantum mechanics [\Wigner,\WeinbergA] has long been known.
In Section II we briefly review Poincar\'e invariance. Following
Weinberg [\WeinbergA] and generalizing the spin 1/2 work of Ryder
[\Ryder], we construct the general boost operator for arbitrary spin
in a form which allows us to produce explicit algebraic expressions
for spinors which describe particles of spin $j$. We present explicit 
expressions for $j=1,\,3/2, ~{\rm and } ~2$. Explicit construction
of covariant spinors for any spin is seen to be reduced to a straightforward 
algebraic exercise. The construction of field operators from the spinors
is noted to be the obvious generalization of the Dirac case.
In Sec. III, using the results of Sec. II,
we derive causal Feynman-Dyson propagators for arbitrary spin.
We note in passing that this propagator, which is the propagator
necessary for perturbative calculations, is not equivalent [\PRCb] to
the Green's function of the Joos--Weinberg equations.
The detailed description of model interactions we leave to future work
since the model or models of the interaction are motivated by the problem
at hand. As is generally the case with phenomenological work, we
choose the model interactions to be the simple Lorentz scalars constructed 
from the field operators
and appropriate kinematical variables.
One of the goals of this work is to learn how data could discriminate
between the possible model couplings and determine their parameters.
In Sec. IV we give conclusions and discuss future applications.
The general philosophy of this work was recently published [\brief]  as
{\it brief report} elsewhere.

\chapter{Construction of the $(j,0)\oplus(0,j)$ Boost Operator}

In this section, we follow the logic of Weinberg [\WeinbergA] and particularly
use a generalization of the spin one--half discussion of Ryder
[\Ryder] to construct the boost operator and the covariant spinors in the
$(j,0)\oplus(0,j)$ representation of the Lorentz group. To set the 
notation and to make this
work self contained, we begin with a brief review of Poincar\'e covariance.
We then briefly summarize the argument [\Wigner,\WeinbergA] that allows
one to construct the boost operators and hence the covariant spinors
directly from invariance principles. We demonstrate how this can be done
in practice by producing explicit expressions for the cases 
$j=1,\,\, 3/2~{\rm and}~2$.

\section{Poincar\'e Transformations}

The Poincar\'e transformations are defined to be the
ten linear and continuous transformations which preserve
$ds^2=dt^2-{d\v x\,}^2=\eta_{\mu\nu} dx^\mu dx^\nu$. We use the metric and,
as far as is possible, the notation of [\Bjorken].
The ten transformations are three rotations about each
of the spacial axes, three boosts along each of the spacial axes, and four 
spacetime translations. These transformations can be summarized by
$$ x^{\prime\mu}={\Lambda^\mu}_\nu x^\nu + a^\mu.\eqn\Lorentz$$
The inertial frame independence of
$ds^2$   requires that the $\Lambda$ matrices satisfy the condition
$$ ~~~{\Lambda^\mu}_\rho
{\Lambda^\nu}_\sigma\,\, \eta_{\mu\nu} =\eta_{\rho\sigma}. \eqn\condition$$ 
In order to exhibit clearly our sign conventions, explicit expressions for
the $\Lambda$ matrices are given in Appendix A.
The transformations $\{\Lambda,a\}$ form a non-abelian group,
$\big[\{\Lambda_1,a_1\},\{\Lambda_2, a_2\}\big]=
\big\{ (\Lambda_1\Lambda_2-\Lambda_2\Lambda_1),(\Lambda_1 a_2-
\Lambda_2 a_1)+(a_1-a_2)\big\}$, with the multiplication law
$\{\overline\Lambda, \overline a\}\{\Lambda, a\}=\{\overline \Lambda\Lambda,
\overline\Lambda a + \overline a\}$, the inverse element
$\{\Lambda,a\}^{-1}=\{\Lambda^{-1},-\Lambda^{-1} a\}$,
and identity element
$\{I,0\}$,
with $I$ a $4\times 4$ identity matrix and $0$ a zero
vector.

For infinitesimal transformations, Eq. \Lorentz ~becomes
$$x^{\prime\mu}=( {\delta^\mu}_\nu + {\lambda^\mu}_\nu )x^\nu + a^\mu, 
\eqn\apag$$
where ${\lambda^\mu}_\nu$ and $a^\mu$ are infinitesimal constants. 
The nonvanishing 
$\lambda^{\mu\nu}={\lambda^\mu}_\epsilon \eta^{\epsilon\nu}$ 
are summarized in Table I.
The ten hermitian generators of the Poincar\'e transformations,
$X_\alpha$ corresponding to the parameter 
$\lambda^\alpha$~~$[\lambda^1=\theta_x,\ldots;
\lambda^4=\varphi_x,\ldots;\lambda^7=a_0,\ldots]$, are defined by
$$X_\alpha \equiv i{\partial x^{\prime~\mu}\over\partial\lambda^\alpha}
\biggr|_{\lambda=0}
{\partial\over\partial x^\mu}~~~~~(\alpha=1,\ldots\ldots,10).
\eqn\apai$$
The three generators of rotation follow from Eqs. (A1--A3) and are the
usual angular momentum operators, $X_{\theta_i}=-L_i,\,\,i=x,\,y,\,z$, 
The three boosts given by Eqs. (A4-A6) yield the 
generators of the Lorentz boosts ($X_{\phi_i}=K_i,\,\,i=x,\,y,\,z$)

$$ K_x
=i\left(t{\partial\over\partial x}+x{\partial\over\partial t}\right),~
 K_y
=i\left(t{\partial\over\partial y}+y{\partial\over\partial t}\right),~
 K_z
=i\left(t{\partial\over\partial z}+z{\partial\over\partial t}\right).\eqn\apbe$$
The  translations given by
$x^{\prime\mu}=x^\mu+a^\mu$
(with $a_\mu$ as  real constant  displacements)
are produced by the four
generators of translations $P_\mu=i{\partial / \partial x^\mu}$.
It should be explicitly noted
that the rotations, boosts and the translations under consideration here
are globally constant. 

By introducing
$$L_{12}=L_z=-L_{21},~~L_{31}=L_y=-L_{13},~~L_{23}=L_x=-L_{32},
~~L_{ij}=\epsilon^{ijk} L_k,
\eqn\apbga$$
$$L_{i0}=-L_{0i}=-K_i, ~~~~~(i=1,2,3),
\eqn\apbgb$$
the algebra associated with these generators can be
summarized as follows
$$[L_{\mu\nu},L_{\rho\sigma}]=i(\eta_{\nu\rho} L_{\mu\sigma}-
\eta_{\mu\rho}L_{\nu\sigma}+\eta_{\mu\sigma}L_{\nu\rho}-
\eta_{\nu\sigma}L_{\mu\rho}),
\eqn\apcb$$
$$
[P_\mu,L_{\rho\sigma}]=i(\eta_{\mu\rho}P_\sigma - \eta_{\mu\sigma}P_\rho),
\eqn\apcc$$
$$
[P_\mu,P_\nu]=0.
\eqn\Algebra$$

\section{Poincar\'e Transformations and Quantum Mechanical States}

The preceding is purely classical physics. What are the implications of 
Poincar\'e invariance for quantum mechanics? This question was answered by
Wigner [\Wigner] and expanded in [\WeinbergA]. We follow these works and 
introduce a quantum mechanical state $\vert state \rangle $.
Let  the \it same \rm system now be observed by another inertial observer
characterised by $\{\Lambda,a\}$. Denote the state as observed by this new
observer by $\vert state \rangle ^\prime$. In order that $\vert state \rangle $
and $\vert state \rangle ^\prime$ be physically acceptable states, they 
must transform as $$\vert state \rangle ^\prime = U(\{\Lambda,a\})~\vert
state \rangle , 
\eqn\states$$
where $U(\{\Lambda,a\})$ is an unitary operator
constrained to satisfy
 $$U(\{\overline\Lambda,\overline a\})
U(\{\Lambda,a\})=U(\{\overline\Lambda\Lambda, \overline\Lambda a + 
\overline a\})..
\eqn\constraint$$
This is simply the requirement that a Poincar\'e transformation
$\{\Lambda,a\}$ followed by $\{\overline
\Lambda,\overline a\}$ has the same effect as the
Poincar\'e transformation  $ \{\overline \Lambda,\overline a\} \{\Lambda,a\}$.
Strictly
speaking \constraint ~is true for infinitesimal transformations.  The finite
Poincar\'e transformations which are constructed by successive application of
infinitesimal transformations will occasionally have a minus sign on the
$r.h.s$ of \constraint. The representation is then said to be a \it
representation up to a sign. \rm This situation will arise when considering
fermionic representations. In such situations the fermionic fields must be so
combined as to yield  observables which are even functions of these fields.
This point is discussed in more detail in Sec.~2.12 of Ref.~[\WeinbergF]. 

Eqs. \states~and \constraint~are sufficient [\Thesis, \WeinbergC] to determine
$U(\{\Lambda, a\})$,
$$U(\{\Lambda, a\})\,=\,
\exp\left[-{i\over 2}\lambda^{\mu\nu} J_{\mu\nu} + i a^\mu P_\mu\right],
\eqn\Unitary$$
where the following algebra is associated with the generators inducing 
the transformation
$$[J_{\mu\nu},J_{\rho\sigma}]=i(\eta_{\nu\rho} J_{\mu\sigma}-
\eta_{\mu\rho}J_{\nu\sigma}+\eta_{\mu\sigma}J_{\nu\rho}-
\eta_{\nu\sigma}J_{\mu\rho}),
\eqn\AlgebraA$$
$$[P_\mu,J_{\rho\sigma}]=i(\eta_{\mu\rho}P_\sigma - \eta_{\mu\sigma}
P_\rho),
\eqn\AlgebraB$$
$$[P_\mu,P_\nu]=0.
\eqn\AlgebraC$$ 

The algebra given by \AlgebraA--\AlgebraC~ 
coincides with the algebra 
associated with the generators of spacetime transformations, \apcb--\Algebra.
This does not imply that
$L_{\mu\nu}$ is necessarily identical to $J_{\mu\nu}$. All that is required is
that both $L_{\mu\nu}$  and $J_{\mu\nu}$ satisfy the same algebra. Even  the
$P_\mu$ appearing in \Unitary~
need not coincide with the generators of spacetime
translations. Specification of a physical state and
determining the effect of a Poincar\'e transformation $\{\Lambda,a\}$ on
that state, therefore, requires an explicit determination of the
generators. 

\section{Spin and Angular Momentum}

As argued by Weinberg [\WeinbergA],
we note that (with the exception of the scalar
field) if one wishes to arrive at the  particle interpretation  within the
framework of Poincar\'e covariant theory of quantum systems, one is forced to
incorporate necessarily non--unitary,  {\it finite}--dimensional  
representations of
the Lorentz group. Since only unitary transformations of physical states allow
for a probabilistic interpretation, the representation spaces of finite
dimensional representations of the Lorentz group cannot be spanned by
``physical states'' defined via \states. The objects which span the finite
dimensional representation spaces are called ``matter fields,''  just
``fields,'' or ``covariant spinors.''  Although the finite--dimensional 
representations are not unitary, they provide the basic ingredients for the
construction of a field theory.

The set of generators $\{\vec J, \vec  K\}$ span a linear vector space 
\footnote{1}{The vector space of the 
generators should not be confused with the vector space $(\equiv \rm
representation~ space)$ on which the generators act.} with
$\vec J$ and $\vec  K$ as the basis vectors. Since the Lorentz
group is  non-compact,  all its finite dimensional
representations are non--unitary.  To
construct these finite dimensional representations, we   explicitly note the
algebra associated  with the Lorentz group \goodbreak 
$$[K_i,J_i]=0,~~i=x,y,z\eqn\xaa$$
$$[J_x,J_y]=i J_z,~~[ K_x, K_y]=-iJ_z, ~~[J_x, K_y]=i
K_z,~~[ K_x,J_y]=i K_z, \eqn\Lorentzalgebra$$ 
and cyclic permutations.  Next we implement the standard rotation
by introducing a new basis:
$$ \vec S_R={1\over 2}(\vec J +i \vec  K),~~
\vec S_L={1\over2}(\vec J-i\vec  K).
\eqn\bpb$$
It follows directly that  $\v S_R$ and $\v S_L$ each satisfy
the algebra of an $SU(2)$ group
$$[(S_R)_i,(S_L)_j]=0,~~~i,j=x,y,z.\eqn\AlbegraRL$$
$$[(S_R)_x, (S_R)_y]=i(S_R)_z,~~[(S_L)_x, (S_L)_y]=i(S_L)_z,\eqn\AlgebraSU$$
and cyclic permutations.
The Lorentz group is thus seen to be essentially equivalent to
$SU(2)_R\otimes SU(2)_L$. 
The irreducible representations of 
$SU(2)_R\otimes SU(2)_L$ are labelled by two numbers $(j_r,j_l)$, ~~
$$\eqalign{(\vec S_R)^2~\phi_{j_r,\sigma_r}=&~j_r(j_r+1)~\phi_{j_r,\sigma_r},
~~(S_R)_z~\phi_{j_r,\sigma_r}=\sigma_r~\phi_{j_r,\sigma_r} \cr
&\sigma_r=j_r,j_r-1,j_r-2,\ldots,-j_r+1,-j_r~~.\cr}
\eqn\Sur$$
$$\eqalign{(\vec S_L)^2~\phi_{j_l,\sigma_l}=&~j_l(j_l+1)~\phi_{j_l,\sigma_l},
~~(S_L)_z~\phi_{j_l,\sigma_l}=\sigma_l~\phi_{j_l,\sigma_l}\cr
&\sigma_l=j_l,j_l-1,j_l-2,\ldots,-j_l+1,-j_l~~.\cr}
\eqn\Sul$$

At this point we will specialize to the $(j,0)$ and $(0,j)$ representations.
These are the simplest representations and are thus a natural place to
start a phenomenology. Results of physical measurements should be independent 
of the representation chosen [\WeinbergQ]. However, arguments of simplicity
enter into building model interactions and simplicity is not always
representation independent.

Under the parity, $(j,0)\leftrightarrow (0,j)$. Thus in order to include
parity, we are led to consider the
$(j,0)\oplus(0,j)$ representation. This also leads to a theory which avoids
any extra degrees of freedom and which naturally incorporates the $2(2j+1)$
spinorial and particle/antiparticle degrees of freedom. We introduce the 
chiral representation
$(j,0)\oplus(0,j)$ covariant spinors
$$\psi_{\smalltype CH}(\v p\,) = \pmatrix{\phi^{^{R}}(\v
p\,)\cr\cr\phi^{^{L}}(\v p\,) }\,\,\,,\eqn\chiral$$ 
where $\phi^{^{R}}(\v p\,)$ represents functions in the $(j,0)$ representation 
space, and 
$\phi^{^{L}}(\v p\,)$ represents functions in the $(0,j)$ representation 
space.

There seems to be some ambiguous statements in the literature [\WeinbergA]
and in textbooks [\Ramond] concerning the hermiticity of the operators
$\v S_R$ and $\v S_L$.
To clarify this we note a basic distinction between the finite
dimensional representation of $\{\v J,\v K\}$ and the infinite dimensional
representations of $\{\v J, \v K\}$. For the $(j,0)$
representation $\v K = -i \v J$, since by definition for the $(j,0)$
representation $\v S_{R}=\v J$ and $\v S_{L}=0$. Similarly for the $(0,j)$
representation $\v K= + i \v J$. As such both $\v
J \pm i\v K$, $\v S_{R}$ and $\v S_{L}$, are hermitian. On the other
hand for the infinite dimensional representations 
both $\v J$ and $\v K$ are, \apai~ and \apbe, are
hermitian. This makes $\v J \pm i\v K$, $\v S_{R}$ and $\v S_{L}$, 
non-hermitian. 
The hermiticity of $\v J\pm i \v K$, and hence 
$\v S_R$ and $\v S_L$,  depends on whether one is concerned with finite
dimensional representations or infinite dimensional representations.

There is an additional difference between the finite dimensional and the
infinite dimensional representations. This difference arises from the
interpretation that spin exists in a separate space, an {\it internal} space. 
The finite dimensional representations thus have spin operators
which commute with the generators of translation, $P_\mu$. This is not true
for the angular momentum operators $L_i$ as can be seen in \apcc.

\section{Construction of the $(j,0)\oplus(0,j)$ Boost Operator}

With this background we are now in a position to construct the
$(j,0)\oplus(0,j)$ boost operator. For a particle at rest in the unprimed 
frame, a Lorentz boost results in a particle with momentum $\v p$. 
The matter fields or covariant spinors transform 
as the physical $\vt state\ra$'s~ (see Eq. \states), 
but with one difference. The
$J_{\mu\nu}$ is replaced by its finite dimensional counterpart and the unitary
operator $U(\{\Lambda,a\})$ is replaced by the non--unitary $D(\{\Lambda,a\})$
which still satisfy the constraint imposed by Poincar\'e covariance,
$$D(\{\overline\Lambda,\overline a\})
D(\{\Lambda,a\})=D(\{\overline\Lambda\Lambda, \overline\Lambda a + \overline
a\}).  \eqn\jbpai$$ 
For the $(j,0)$ and $(0,j)$ representations, we obtain
$$\phi^{^R}(\vec p\,) = \exp[\vec J\cdot\vec\varphi]~~\phi^{^R}(\vec 0) 
\eqn\jbpbo$$
$$\phi^{^L}(\vec p\,) = \exp[-\vec J\cdot\vec\varphi]~~\phi^{^L}(\vec 0),
\eqn\jbpba$$
where the boost  parameter $\v \varphi$ is defined as 
$$\cosh(\varphi\,)\,=\,\gamma\,=\,{1\over\sqrt{1-v^2}}\,=\,{E\over m},~~
\sinh( \varphi\,)\,=\,v\gamma\,=\,{\vt \v p\, \vt\over m},~~\hat\varphi={\v p
\over\vt \v p\,\vt},
\eqn\parameter$$
with $\v p$ the three-momentum of the particle .

As a consequence, the chiral representation
\footnote{2}{We call this representation the ``chiral'' representation
because for $j=1/2$ the representation coincides with the chiral
representation of the Dirac spin one--half formalism.}
$(j,0)\oplus(0,j)$ relativistic
covariant spinors defined by Eq. \chiral~ transform as
$$\psi_{\smalltype CH}(\v p\,) = \pmatrix{\exp(\vec J\cdot\vec\varphi)&0\cr\cr
                                0&\exp(-\vec J\cdot\vec\varphi)}
\psi_{\smalltype CH}(\v 0) .\eqn\Chiralboost$$

\medskip
We also introduce a canonical representation which for spin one--half
is equivalent to the canonical representation used in [\Bjorken].
The transformation matrix A which relates these representations is given by
$$\psi_{{\smalltype CA}}(\v p\,) = A~ \psi_{\smalltype CH}(\v
p\,),~~A={1\over\sqrt 2}\pmatrix {I&I\cr 
                                                             I&-I\cr}.
\eqn\A$$
Each  entry $I$ in the matrix $A$ represents a $(2j+1)\times (2j+1)$
identity matrix.

In the canonical representation,the  covariant spinors are 
$$\psi_{{\smalltype CA}}(\vec p\,) = {1\over \sqrt 2}\pmatrix
{\phi^{^{\smalltype R}}(\vec p\,) +\phi^{^{\smalltype L}}(\vec p\,)\cr\cr
 \phi^{^{\smalltype R}}(\vec p\,) - \phi^{^{\smalltype L}}(\vec p\,)\cr},
\eqn\Canonical$$
with the even and odd under parity components of the spinors
separated as the upper and lower components, respectively.

Referring to Eq. \Chiralboost, we identify
the chiral representation boost matrix as
$$M_{\smalltype CH}(\v p\,)=\pmatrix{\exp(\vec J\cdot\vec\varphi)&0\cr\cr
                                0&\exp(-\vec J\cdot\vec\varphi)}.
\eqn\jbpbe$$
The boost matrix in the canonical representation reads
$$M_{{\smalltype CA}}(\vec p\,) =
\pmatrix{\cosh(\vec J\cdot\vec\varphi) & \sinh (\vec J\cdot\vec\varphi)\cr\cr
\sinh (\vec J\cdot\vec \varphi) &  \cosh(\vec J\cdot\vec\varphi)\cr}.
\eqn\Canonicalboost$$
If $\v J$ is set equal to $\v \sigma/2$ the boost matrix given by Eq.
\Canonicalboost~ coincides
with the boost for Dirac spinors in the standard Bjorken and Drell [\Bjorken]
representation. 
$M_{{\smalltype CA}}(\vec p\,) $ contains all the information needed to 
construct any $(j,0)\oplus (0,j)$ relativistic covariant spinor. 
In the next section, we provide the explicit results for $j=~$1, 3/2 and 2.
The examples are chosen not only to
demonstrate the procedure of constructing the arbitrary--spin covariant spinors,
for mesons and baryons, but also to provide readily available covariant spinors
through spin two. Elsewhere [\Threehalf], we use the spin-$3/2$ covariant
spinors obtained here to study the scattering of a spin-$3/2$ baryon from an
external Coulomb field. 

\chapter{ $(j,0)+(0,j)$ Covariant Spinors}

\section{  $(1,0)\oplus(0,1)$ Covariant Spinors}

The representation space of the $(1,0)\oplus(0,1)$ covariant spinors
is a six
dimensional internal space. The basis vectors for a particle at rest 
can be chosen to be, in the canonical representation, 
\medskip
$$u_{_{\smalltype +1}}(\v 0) = \pmatrix{m\cr
                                       0\cr
                                       0\cr
                                       0\cr
                                       0\cr
                                       0\cr},~~
u_{_{\smalltype 0}}(\v 0) = \pmatrix{   0\cr
                                       m\cr
                                       0\cr
                                       0\cr
                                       0\cr
                                       0\cr},~~
u_{_{\smalltype -1}}(\v 0) = \pmatrix{0\cr
                                       0\cr
                                       m\cr
                                       0\cr
                                       0\cr
                                       0\cr},$$
$$v_{_{\smalltype +1}}(\v 0) = \pmatrix{0\cr
                                       0\cr
                                       0\cr
                                       m\cr
                                       0\cr
                                       0\cr},~~
v_{_{\smalltype 0}}(\v 0) = \pmatrix{   0\cr
                                       0\cr
                                       0\cr
                                       0\cr
                                       m\cr
                                       0\cr},~~
v_{_{\smalltype -1}}(\v 0) = \pmatrix{0\cr
                                       0\cr
                                       0\cr
                                       0\cr
                                       0\cr
                                       m\cr}.\eqn\jbpbg$$
The  norm is chosen for convenience in considering
the $m\rightarrow 0$ limit. This choice of the basis vectors, and the
interpretation attached to them that $u_{\sigma}(\v 0)$ represents a particle at
rest with the z-component of its spin to be $\sigma$ $(\sigma=0,\pm 1)$ and
$v_{\sigma}(\v 0)$ an antiparticle at rest
with the z-component of its spin to be $\sigma$
requires that  $J_z$ be diagonal. This gives [\Schiff],
$$J_x={1\over\sqrt 2}\pmatrix
{0&1&0\cr
1&0&1\cr
0&1&0\cr},~~
J_y={1\over\sqrt 2}\pmatrix
{0&-i&0\cr
i&0&-i\cr
0&i&0\cr},~~
J_z=\pmatrix{1&0&0\cr 0&0&0 \cr 0&0&-1\cr}.\eqn\jbpbi$$
The boost matrix $M_{{\smalltype CA}}(\vec p\,)$ takes the 
covariant spinor of a
particle at rest, $\psi_{{\smalltype CA}}(\vec 0)$, to 
$\psi_{{\smalltype CA}}(\vec p\,)$, the  
covariant spinor of the same particle
with momentum $\vec p$
$$\psi_{_{ CA}}(\vec p\,)= M_{_{ CA}}(\vec p\,)~\psi_{_{ CA}}(\vec 0).
\eqn\jbpbi$$ 
The $\cosh(\v J\cdot\v\varphi)$ which appears in the  covariant spinor
boost matrix  \footnote{3}{See  Appendix B for the general 
expansions of $\cosh(\v J\cdot\v \varphi)$ and  $\sinh(\v J\cdot\v\varphi)$ } 
\Canonicalboost ~can be expanded to yield 
$$\cosh(\vec J\cdot\vec \varphi)=
\cosh\left( 2\vec J\cdot{\vec\varphi\over 2}\right)
=1+2(\vec J\cdot\hat p\,)(\vec J\cdot\hat p\,)\sinh^2\left({\varphi\over 2}
\,\,\,. \right)\eqn\jbpbh$$
Note that
$$\sinh \left({\varphi\over2}\right)= \left({E-m\over 2m}\right)^{1\over2},
\eqn\jbpca $$
and
$$\vec J\cdot\hat p ={1\over\vert \vec p\vert}\vec J\cdot\vec p
={1\over{(E^2-m^2)^{1\over 2}}}
\left( J_x p_x + J_y p_y + J_z p_z\right).
\eqn\jbpcb$$
Substituting for $J_i$ from \jbpbi ~gives the matrix $\v J\cdot\hat p$
$$\vec J\cdot\hat p =
{1\over {(E^2-m^2)^{1\over 2} } }\pmatrix {
p_z&{1\over\sqrt 2}(p_x-ip_y)&0\cr\cr
{1\over\sqrt 2}(p_x+ip_y)&0&{1\over\sqrt 2}(p_x-ip_y)\cr\cr
0&{1\over\sqrt 2}(p_x+ip_y)&-p_z\cr}.
\eqn\jbpcc$$
This, when substituted into Eq.. \jbpbh, gives
$$\cosh(\vec J\cdot \vec \varphi)=1+
{1\over m(E+m)}\pmatrix{p_z^2+{1\over 2}p_{_+} p_{_-} & {1\over \sqrt 2}p_z 
p_{_-}&
{1\over 2}p_{_-}^2\cr\cr
{1\over \sqrt 2}p_z p_{_+} & p_{_+} p_{_-} &-{1\over \sqrt 2}p_z p_{_-}\cr\cr
{1\over 2}p_{_+}^2 &-{1\over \sqrt 2}p_z p_{_+} &
p_z^2+{1\over 2}p_{_+} p_{_-} \cr},
\eqn\jbpce$$
where
$$p_{_\pm} \equiv p_x \pm i p_y,\eqn\jbpcd$$
Similarly $\sinh(\v J\cdot\v\varphi)$ which appears in the canonical
representation boost matrix
for the covariant spinors \Canonicalboost ~can be expanded as 
$$\sinh(\v J\cdot\v\varphi)=
\sinh\left(2\v J\cdot{\v \varphi\over 2}\right) =
2(\v J\cdot\hat p\,)\cosh\left({\varphi\over 2}\right)
\sinh\left({\varphi\over 2}\right).
\eqn\jbpcf$$
Using Eq. \jbpcc~
and noting that 
$$\cosh\left({\varphi\over 2}\right)=
\left({E+m\over 2m}\right)^{1\over 2}.\eqn\jbpcg$$
yields
$$\sinh(\vec J\cdot\vec \varphi)=
{1\over m} \pmatrix{ p_z & {1\over\sqrt 2} p_{_-} & 0\cr\cr
{1\over\sqrt 2} p_{_+} & 0 &{1\over\sqrt 2} p_{_-}\cr\cr
0 &{1\over\sqrt 2} p_{_+} & -p_z}.\eqn\jbpch$$

Substituting $\sinh(\vec J\cdot\vec \varphi)$ and $\cosh(\vec J\cdot\vec 
\varphi)$
into \Canonicalboost ~provides an explicit expression
for the canonical representation boost operator for the $(1,0)\oplus(0,1)$ 
covariant spinors.
Applying the boost operator \Canonicalboost~
to the rest spinors \jbpbg~ and utilizing the identities
\jbpce~and\jbpch~yields the
 $(1,0)\oplus(0,1)$ covariant spinors
$$  u_{_ {+1}}(\vec p\,)=
\pmatrix{m+\left[(2p_z^2~+~p_{_{+}} p_{_{-}}) / 2(E+m)\right]\cr\cr
                      {p_z p_{_{+}}/{\sqrt 2}(E+m)}\cr\cr
              { p_{_{+}}^2/ 2(E+m) }\cr\cr
               p_z\cr\cr
                   {p_{_{+}}/{\sqrt 2}}\cr\cr
                   0\cr},\eqn\jbpcia$$

\endpage
$$u_{_{0}}(\vec p\,)=\pmatrix{{p_z p_{_{-}}/{\sqrt 2}(E+m)}\cr\cr
                      m+\left[{p_{_{+}} p_{_{-}}/(E+m) }\right]\cr\cr
                       -{p_z p_{_{+}}/{\sqrt 2}(E+m)}\cr\cr
                       {p_{_{-}}/{\sqrt 2}}\cr\cr
                          0\cr\cr
                        {p_{_{+}}/{\sqrt 2}}\cr},\eqn\jbpcib$$

$$u_{_{-1}}(\vec p\,)=\pmatrix{ { p_{_{-}}^2/ 2(E+m) }\cr\cr
                             -{p_z p_{_{-}}/{\sqrt 2}(E+m)}\cr\cr
                   m+\left[{(2p_z^2~+~p_{_{+}} p_{_{-}})/ 2(E+m)}\right]\cr\cr
                      0\cr\cr
                      {p_{_{-}}/{\sqrt 2}}\cr\cr
                   -p_z\cr},\eqn\jbpcic$$

\endpage
$$v_{_{+1}}(\vec p\,)=\pmatrix{ p_z\cr\cr
                   {p_{_{+}}/{\sqrt 2}}\cr\cr
                   0\cr\cr
        m+\left[{(2p_z^2~+~p_{_{+}} p_{_{-}})/ 2(E+m)}\right]\cr\cr
                      {p_z p_{_{+}}/{\sqrt 2}(E+m)}\cr\cr
              { p_{_{+}}^2/ 2(E+m) }\cr},\eqn\jbpcid$$

$$
v_{_{0}}(\vec p\,)= \pmatrix{ {p_{_{-}}/{\sqrt 2}}\cr\cr
                          0\cr\cr
                        {p_{_{+}}/{\sqrt 2}}\cr\cr
                 {p_z p_{_{-}}/{\sqrt 2}(E+m)}\cr\cr
                      m+\left[{p_{_{+}} p_{_{-}}/(E+m) }\right]\cr\cr
                       -{p_z p_{_{+}}/{\sqrt 2}(E+m)}\cr},
\eqn\jbpcie$$

\endpage
$$v_{_{-1}}(\vec p\,)= \pmatrix{0\cr\cr
                      {p_{_{-}}/{\sqrt 2}}\cr\cr
                   -p_z\cr\cr
             { p_{_{-}}^2/ 2(E+m) }\cr\cr
                             -{p_z p_{_{-}}/{\sqrt 2}(E+m)}\cr\cr
                   m + \left[{(2p_z^2~+~p_{_{+}} p_{_{-}})/ 2(E+m)}\right]\cr} .
\eqn\jbpcif$$

It is important to note that the \mo~limit of these spinors is well--defined
and physical. Consider a massless particle travelling along the $\hat z$ axis 
(for an arbitrary direction
the quantization axis for the angular momentum would have to be chosen 
accordingly). 
For this case only $u_{\pm 1}(\v p\,)$ and $v_{\pm 1}(\v p\,)$ survive
while 
$u_{ 0}(\v p\,)$ and $v_{ 0}(\v p\,)$ 
vanish identically. 
Explicitly the \mo~limits of the covariant spinors are
$$\lim_{m\rightarrow 0} u_{+{1}}( p_z\,) =
\pmatrix{E\cr 0\cr 0\cr 
         E\cr 0\cr 0\cr},~
\lim_{m\rightarrow 0} u_{{0}}( p_z\,) =
\pmatrix{0\cr 0\cr 0\cr 
         0\cr 0\cr 0\cr },~
\lim_{m\rightarrow 0} u_{-{1}}( p_z\,) =
\pmatrix{0\cr 0\cr E\cr
         0\cr 0\cr -E\cr},\eqn\jxape$$
and
$$\lim_{m\rightarrow 0} v_{+{1}}( p_z\,) =
\pmatrix{E\cr 0\cr 0\cr 
         E\cr 0\cr 0\cr},~
\lim_{m\rightarrow 0} v_{{0}}( p_z\,) =
\pmatrix{0\cr 0\cr 0\cr 
         0\cr 0\cr 0\cr },~
\lim_{m\rightarrow 0} u_{-{1}}( p_z\,) =
\pmatrix{0\cr 0\cr -E\cr
         0\cr 0\cr E\cr}.\eqn\jxapf$$
While $u_{+{1}}(p_z)$ and $v_{+{1}}(p_z)$ are identical
the $u_{-{1}}(p_z)$ $v_{-{1}}(p_z)$ differ by a relative minus sign. As one 
would wish, only the $\vert m\vert=j$ spinors remain non--zero in the
\mo~limit.

\section{$(3/2,0)\oplus(0,3/2)$ Covariant Spinors}

A covariant description of spin 3/2 baryons, such as the
$\Delta$(1232), can be based on the $(3/2,0)\oplus(0,3/2)$ covariant spinors.
As for all the $(j,0)\oplus(0,j)$
spinors, the covariant spinors for spin $3/2$ have 
exactly the correct number of spinorial and particle/antiparticle degrees of
freedom.The canonical representation $(3/2)\oplus(0,3/2)$ covariant spinors
are obtained in a similar fashion as were the $(1,0)\oplus(0,1)$ covariant 
spinors in the last section. 

We first note that Eqs. (B3) and (B4) for $j=3/2$ give
$$\cosh(2\v J\cdot\v \varphi)=\cosh\varphi\Big[I+{1\over 2} 
\big\{(2\v J\cdot\hat p\,)^2-I\big\}\sinh^2\varphi\Big],
\eqn\jbpde$$
$$\sinh(2\v J\cdot\v \varphi)=(2\v J\cdot\hat p\,)\sinh\varphi\Big[I+{1\over 6} 
\big\{(2\v J\cdot\hat p\,)^2-I\big\}\sinh^2\varphi\Big],
\eqn\jbpdf$$
Using \jbpca, \jbpcb ~ and \jbpcg~ gives
$$\cosh(\v J\cdot\v\varphi) =\left({E+m\over 2m}\right)^{1/2}
\left[I+{1\over 2}\left\{ {(2\v J\cdot\v p\,)^2\over
(E^2-m^2)}-I\right\}\left({E-m\over 2m} 
\right)\right]
\eqn\jbpea$$
$$\eqalign{&\sinh(\v J\cdot\v\varphi) =\cr
&\left({E+m\over 2m}\right)^{1/2} 
\left[{2\v J\cdot \v p\over (E+m)}+{1\over 6} {2\v J\cdot \v p\over (E+m)}
\left\{ {(2\v J\cdot\v p\,)^2\over (E^2-m^2)}-I\right\}\left({E-m\over 2m}
\right)\right].\cr}
\eqn\jbpeb$$
These expansions when substituted in Eq. \Canonicalboost ~provide the 
canonical representation boost for the
$(3/2,0)\oplus(0,3/2)$ covariant spinors.

For the rest spinors we chose the norm such that in the $m\rightarrow 0$ 
limit the
rest spinors vanish and the boosted spinors have a
non--singular norm. The simplest choice for $u_\sigma(\v 0)$ and $v_\sigma(\v
0)$ are eight--element column vectors each 
with  a single entry of $m^{3/2}$ in the
appropriate row and zero elsewhere. 
Interpreting these rest spinors  as eigenstates of $J_z$ requires
a representation in which $J_z$ is diagonal,
$$J_x={1\over 2} \pmatrix{ 0 & \sqrt 3 & 0 & 0\cr
                           \sqrt 3 & 0 & 2 & 0 \cr
                           0 & 2 & 0 & \sqrt 3 \cr
                           0 & 0 & \sqrt 3 & 0\cr},
J_y={1\over 2} \pmatrix{ 0 & -i \sqrt 3 & 0 & 0\cr
                           i\sqrt 3 & 0 & -2i &0 \cr
                           0 & 2i & 0 & -i\sqrt 3 \cr
                           0 & 0 & i\sqrt 3 & 0\cr},$$
$$J_z={1\over 2} \pmatrix{   3 & 0 & 0 & 0\cr
                           0 & 1 & 0 & 0 \cr
                           0 & 0 & -1 & 0  \cr
                           0 & 0 & 0 & -3 \cr}. 
\eqn\jbpee $$
Substituting these into \jbpea~and \jbpeb, and putting the result into
\Canonicalboost~ provides an explicit expression
for the
$(3/2,0)\oplus(0,3/2)$ boost operator in the canonical representation, which
when applied to the rest spinors, gives
\endpage
$$\eqalign{ u_{ +{3\over 2}}(\v p\,) =&  {m^{1\over 2}} 
\left({E+m\over 2m}\right)^{1\over 2}\cr
&\times\pmatrix{ { (9p_z^2 + 3p_{_{+}}p_{_{-}} + 5m^2 + 4Em -E^2)/ {4(m+E)}
}\cr\cr\cr 
{    {\sqrt{3}} p_{_{+}} p_z / (m+E)    }\cr\cr\cr
{\sqrt{3} p_{_{+}}^2 / 2(m+E)}\cr\cr\cr
0\cr\cr\cr
{ p_z(9 p_z^2 + 7 p_{_{+}} p_{_{-}} + 13 m^2 +12Em -E^2)/ 4(m+E)^2 }\cr\cr\cr
{ {\sqrt{3}} p_{_{+}}
(13 p_z^2 + 7 p_{_{+}} p_{_{-}} + 13 m^2 +12Em -E^2)/ 12(m+E)^2 }\cr\cr\cr
{ {\sqrt{3}} p_{_{+}}^2 p_z / 2(m+E)^2 }\cr\cr\cr
{ p_{_{+}}^3/ 2(m+E)^2}\cr},\cr}
\eqn\jbpefa$$
\endpage
$$\eqalign{ u_{ +{1\over 2}}(\v p\,) =&  {m^{1\over 2}} 
\left({E+m\over 2m}\right)^{1\over 2}\cr
&\times\pmatrix{ 
{\sqrt{3}} p_{_{-}} p_z / (m+E)    \cr\cr\cr
(p_z^2 + 7 p_{_{+}} p_{_{-}} + 5 m^2 +4Em -E^2)/ 4(m+E) \cr\cr\cr
0\cr\cr\cr
{\sqrt{3} }p_{_{+}}^2 / 2(m+E)\cr\cr\cr
{\sqrt{3}} p_{_{-}}(13 p_z^2 + 7 p_{_{+}} p_{_{-}} + 13 m^2 +12Em -E^2)/
12(m+E)^2 \cr\cr\cr 
p_z (p_z^2 + 19p_{_{+}}p_{_{-}} + 13m^2 + 12Em -E^2)/ {12(m+E)^2} \cr\cr\cr
p_{_{+}}(p_z^2 + 10 p_{_{+}} p_{_{-}} + 13 m^2 +12Em -E^2)/ 6(m+E)^2 \cr\cr\cr
- {\sqrt{3}} p_{_{+}}^2 p_z / 2(m+E)^2 \cr}},
\eqn\jbpefb$$
\endpage
$$\eqalign{ u_{ -{1\over 2}}(\v p\,) =&  {m^{1\over 2}} 
\left({E+m\over 2m}\right)^{1\over 2}\cr
&\times\pmatrix{ 
{\sqrt{3}} p_{_{-}}^2 / 2(m+E)    \cr\cr\cr
0\cr\cr\cr
(p_z^2 + 7 p_{_{+}} p_{_{-}} + 5 m^2 +4Em -E^2)/ 4(m+E) \cr\cr\cr
-{\sqrt{3} }p_{_{+}} p_z / (m+E)\cr\cr\cr
 {\sqrt{3}} p_{_{-}}^2 p_z / 2(m+E)^2 \cr\cr\cr
p_{_{-}}(p_z^2 + 10 p_{_{+}} p_{_{-}} + 13 m^2 +12Em -E^2)/ 6(m+E)^2 \cr\cr\cr
-p_z (p_z^2 + 19p_{_{+}}p_{_{-}} + 13m^2 + 12Em -E^2)/ {12(m+E)^2} \cr\cr\cr
{\sqrt{3}} p_{_{+}}(13 p_z^2 + 7 p_{_{+}} p_{_{-}} + 13 m^2 +12Em -E^2)/ 
12(m+E)^2 \cr}},
\eqn\jbpefc$$
\endpage
$$\eqalign{ u_{ -{3\over 2}}(\v p\,) =&  {m^{1\over 2}} 
\left({E+m\over 2m}\right)^{1\over 2}\cr
&\times\pmatrix{ 
0\cr\cr\cr
\sqrt{3} p_{_{-}}^2 / 2(m+E)\cr\cr\cr
-{\sqrt{3}} p_{_{-}} p_z / (m+E) \cr\cr\cr
{ (9p_z^2 + 3p_{_{+}}p_{_{-}} + 5m^2 + 4Em -E^2)/ {4(m+E)} }\cr\cr\cr
 p_{_{-}}^3/ 2(m+E)^2\cr\cr\cr
- {\sqrt{3}} p_{_{-}}^2 p_z / 2(m+E)^2 \cr\cr\cr
 {\sqrt{3}} p_{_{-}}
(13 p_z^2 + 7 p_{_{+}} p_{_{-}} + 13 m^2 +12Em -E^2)/ 12(m+E)^2 \cr\cr\cr
- p_z(9 p_z^2 + 7 p_{_{+}} p_{_{-}} + 13 m^2 +12Em -E^2)/ 4(m+E)^2 
        \cr}}.
\eqn\jbpefd$$
An inspection of the boost given by Eq. \Canonicalboost
 ~immediately reveals  that the four
$v_\sigma(\v p\,)'s$ can now be readily obtained by interchanging
 the four bottom
elements with the four top elements of the respective $u_\sigma(\v p\,)'s$,
i.e.
$$v_\sigma(\v p\,) = \gamma_5\,\,u_\sigma(\v p\,), 
\eqn\jbpeh$$ 
where $\gamma_5$ is
$$\gamma_5=\pmatrix{0&I\cr I&0},
\eqn\jbpei$$
with $I$ the $(2j+1)\times(2j+1)$ identity matrix.

It can be readily verified
that for a massless particle traveling along the $\hat z$ axis, only
the $u_{\pm {3\over 2}}(\v p\,)$ and $v_{\pm {3\over 2}}(\v p\,)$ survive.
The $u_{\pm {1\over 2}}(\v p\,)$ and $v_{\pm {1\over 2}}(\v p\,)$  vanish
identically. Explicitly, the non--zero spinors are given by
$$\lim_{m\rightarrow 0} u_{+{3\over 2}}( p_z\,) ={\sqrt 2} E^{3\over 2}
\pmatrix{1\cr 0\cr 0\cr 0\cr
         1\cr 0\cr 0\cr 0\cr},~~~~
\lim_{m\rightarrow 0} u_{-{3\over 2}}( p_z\,) ={\sqrt 2} E^{3\over 2}
\pmatrix{0\cr 0\cr 0\cr 1\cr
         0\cr 0\cr 0\cr -1\cr},
\eqn\jbpfhcd$$
and
$$\lim_{m\rightarrow 0} v_{+{3\over 2}}( p_z\,) ={\sqrt 2} E^{3\over 2}
\pmatrix{1\cr 0\cr 0\cr 0\cr
         1\cr 0\cr 0\cr 0\cr},~~~~
\lim_{m\rightarrow 0} v_{-{3\over 2}}( p_z\,) ={\sqrt 2} E^{3\over 2}
\pmatrix{0\cr 0\cr 0\cr -1\cr
         0\cr 0\cr 0\cr 1\cr}.
\eqn\jbpgocd$$
While $u_{+{3\over 2}}(p_z)$ and $v_{+{3\over 2}}(p_z)$ are identical
the $u_{-{3\over 2}}(p_z)$ $v_{-{3\over 2}}(p_z)$ again differ by a minus sign.

For a detailed analysis of certain kinematic acausality [\brief] in the
massless limit of  Weinberg's work [\WeinbergB]  and its resolution the reader
is referred to one of our recent work [\Paradox].

\section{$(2,0)\oplus(0,2)$ Covariant Spinors }

In order to calculate the $(2,0)\oplus(0,2)$ covariant spinors we
follow the now familiar procedure. The boost is given by \Canonicalboost~
with
$$\cosh(\v J\cdot\v \varphi) = I + {(\v J\cdot \v p\,)^2\over m(m+E)}
+{1\over 6}
{(\v J\cdot \v p\,)^2((\v J\cdot \v p\,)^2-\v p^{~2})\over m^2(m+E)^2} 
\eqn\jbpgda$$
$$\sinh(\v J\cdot\v \varphi) =  {\v J\cdot \v p\over m}
+{1\over 3}
{\v J\cdot \v p~((\v J\cdot \v p\,)^2-\v p^{~2})\over m^2(m+E)} 
\eqn\jbpgdb$$
and
$$J_x = \pmatrix{ 0 & 1 & 0 & 0 & 0\cr 
                  1 & 0 & \sqrt{3/2}  & 0 & 0\cr
                  0 &\sqrt{3/2}& 0 & \sqrt{3/2}  & 0\cr
                  0 & 0 & \sqrt{3/2}  & 0 & 1\cr
                  0 & 0 & 0 & 1 & 0\cr},$$
$$J_y = \pmatrix{ 0 & -i & 0 & 0 & 0\cr 
                  i & 0 & -i\sqrt{3/2}   & 0 & 0\cr
                  0 &i\sqrt{3/2}& 0 & -i\sqrt{3/2}  & 0\cr
                  0 & 0 & i\sqrt{3/2}  & 0 & -i\cr
                  0 & 0 & 0 & i & 0\cr},$$
$$ J_z =\pmatrix{2 & 0 & 0 & 0 & 0\cr
                 0 & 1 & 0 & 0 & 0\cr
                 0 & 0 & 0 & 0 & 0\cr
                 0 & 0 & 0 & -1 & 0\cr
                 0 & 0 & 0 & 0 & -2\cr}.
\eqn\jbpgc$$
Application of this boost to rest spinors $u_\sigma(\v 0)$ and
$v_\sigma(\v 0)$, each in the form of column vectors with a 
single entry of $m^2$
in the
appropriate row and zero elsewhere, yields the following $(2,0)\oplus(0,2)$
covariant spinors
\vfill\eject
\vbox{
$$u_{+2}(\v p\,)=\pmatrix{ {(}8 p_{z}^4+8( p_{_{-}} p_{_{+}} + 2 m^2 + 2 E m)
p_{z}^2 + p_{_{-}}^2 p_{_{+}}^2 {+4m( m + E ) p_{_{-}} p_{_{+}}}\cr 
{+4 m^2(m+E)^2}{)} /4 (m+E)^2 \cr\cr\cr 
(4 p_{_{+}} p_{z}^3 +(3 p_{_{-}} p_{_{+}}^2
+6m(m+E)p_{_{+}})p_{z})/2(m+E)^2\cr\cr\cr 
{\sqrt{6}} (2 p_{_{+}}^2 p_{z}^2 +
p_{_{-}} p_{_{+}}^3 + 2 m(m+E) p_{_{+}}^2)/4(m+E)^2\cr\cr\cr
p_{_{+}}^3 p_{z}/2(m+E)^2\cr\cr\cr 
p_{_{+}}^4/4(m+E)^2\cr\cr\cr (2 p_{z}^3 +
(p_{_{-}} p_{_{+}} +2 m^2 +2 Em) p_{z})/(m+E)\cr\cr\cr
 ( 4 p_{_{+}} p_{z}^2 +
p_{_{-}} p_{_{+}}^2 + 2m(m+E) p_{_{+}})/2(m+E)\cr\cr\cr
{\sqrt{6}}~ p_{_{+}}^2
p_{z}/ 2(m+E)\cr\cr\cr p_{_{+}}^3/2(m+E)\cr\cr\cr 0\cr}
\eqn\jbpgfa$$ 
}
\endpage
\vbox{
$$u_{+1}(\v p\,)=\pmatrix{(4 p_{_{-}} p_{z}^3 +3 (p_{_{-}}^2 p_{_{+}} + 2m(m+E)
p_{_{-}})p_{z})/2(m+E)^2\cr\cr\cr
 {(}2(2 p_{_{-}} p_{_{+}} + m^2 +E m) p_{z}^2 + 2
p_{_{-}}^2 p_{_{+}}^2 +5m (m+E) p_{_{-}} p_{_{+}} \cr +2m^2 (m+E)^2
{)}/2(m+E)^2\cr\cr\cr 
{\sqrt{6}}(p_{_{-}} p_{_{+}}^2 +m(m+E) p_{_{+}})
p_{z}/2(m+E)^2\cr\cr\cr 
( 2 p_{_{-}} p_{_{+}}^3 +3m(m+E) p_{_{+}}^2) /
2(m+E)^2\cr\cr\cr 
-p_{_{+}}^3 p_{z}/2(m+E)^2\cr\cr\cr 
(4 p_{_{-}} p_{z}^2 +
p_{_{-}}^2 p_{_{+}} + 2m(m+E) p_{_{-}})/2(m+E)\cr\cr\cr 
(2 p_{_{-}} p_{_{+}} +
m(m+E)) p_{z}/ (m+E)\cr\cr\cr 
{\sqrt{6}}(p_{_{-}} p_{_{+}}^2 +
m(m+E)p_{_{+}})/2(m+E)\cr\cr\cr 
0\cr\cr\cr 
p_{_{+}}^3/2(m+E)\cr}
\eqn\jbpgfb$$ 
}
\endpage
\vbox{
$$u_{0}(\v p\,)=\pmatrix{{\sqrt{6}}(2 p_{_{-}}^2 p_{z}^2 + p_{_{-}}^3 p_{_{+}} +
2m(m+E)p_{_{-}}^2)/4(m+E)^2\cr\cr\cr 
{\sqrt{6}}(p_{_{-}}^2 p_{_{+}}
+m(m+E)p_{_{-}})p_{z}/2(m+E)^2\cr\cr\cr 
(3 p_{_{-}}^2 p_{_{+}}^2 + 6m(m+E)p_{_{-}}
p_{_{+}} + 2m^2(m+E)^2)/2(m+E)^2\cr\cr\cr 
-{\sqrt{6}}(p_{_{-}} p_{_{+}}^2 +
m(m+E)p_{_{+}}) p_{z}/2(m+E)^2\cr\cr\cr 
{\sqrt{6}}(2 p_{_{+}}^2 p_{z}^2 + p_{_{-}}
p_{_{+}}^3 +2m(m+E)p_{_{+}}^2)/4(m+E)^2\cr\cr\cr 
{\sqrt{6}}p_{_{-}}^2
p_{z}/2(m+E)\cr\cr\cr 
{\sqrt{6}}(p_{_{-}}^2 p_{_{+}} + m(m+E)p_{_{-}})/2(m+E)\cr\cr\cr
0\cr\cr\cr 
{\sqrt{6}}(p_{_{-}} p_{_{+}}^2+ m(m+E)p_{_{+}})/2(m+E)\cr\cr\cr
-{\sqrt{6}}p_{_{+}}^2 p_{z}/2(m+E)\cr}
\eqn\jbpgfc$$ 
}
\endpage
\vbox{
$$u_{-1}(\v p\,)=\pmatrix{ p_{_{-}}^3 p_{z}/2(m+E)^2\cr\cr\cr 
(2 p_{_{-}}^3 p_{_{+}}
+3m(m+E) p_{_{-}}^2)/2(m+E)^2\cr\cr\cr 
-{\sqrt{6}}(p_{_{-}}^2 p_{_{+}} +m(m+E)p_{_{-}})
p_{z}/2(m+E)^2\cr\cr\cr 
{(} 2(2 p_{_{-}} p_{_{+}} + m^2 + E m)p_{z}^2 + 2
p_{_{-}}^2 p_{_{+}}^2 + 5m(m+E) p_{_{-}} p_{_{+}} \cr
+2m^2(m+E)^2{)}/2(m+E)^2\cr\cr\cr 
-(4 p_{_{+}} p_{z}^3 +3( p_{_{-}} p_{_{+}}^2 +
2m(m+E) p_{_{+}}) p_{z}/2(m+E)^2\cr\cr\cr 
p_{_{-}}^3/2(m+E)\cr\cr\cr 
0\cr\cr\cr
{\sqrt{6}}(p_{_{-}}^2 p_{_{+}} +m(m+E)p_{_{-}})/2(m+E)\cr\cr\cr 
-(2 p_{_{-}}
p_{_{+}} + m(m+E))p_{z}/(m+E)\cr\cr\cr 
(4 p_{_{+}} p_{z}^2 + p_{_{-}} p_{_{+}}^2 +
2m(m+E) p_{_{+}})/2(m+E)\cr}
\eqn\jbpgfd$$ 
}
\vfill\eject
\vbox{
$$u_{-2}(\v p\,)=\pmatrix{p_{_{-}}^4/4(m+E)^2\cr\cr\cr 
-p_{_{-}}^3
p_{z}/2(m+E)^2\cr\cr\cr 
{\sqrt{6}}(2 p_{_{-}}^2 p_{z}^2 + p_{_{-}}^3 p_{_{+}}
+2m(m+E)p_{_{-}}^2)/4(m+E)^2\cr\cr\cr 
-(4 p_{_{-}} p_{z}^3 + 3(p_{_{-}}^2 p_{_{+}}
+ 2m(m+E)p_{_{-}}) p_{z})/2(m+E)^2\cr\cr\cr 
{(} 8p_{z}^4 + 8 (p_{_{-}} p_{_{+}} +
2m(m+E))p_{z}^2 + p_{_{-}}^2 p_{_{+}}^2 +4m(m+E)p_{_{-}} p_{_{+}} \cr
+4m^2(m+E)^2{)}/4(m+E)^2\cr\cr\cr 
0\cr\cr\cr 
p_{_{-}}^3/2(m+E)\cr\cr\cr
-{\sqrt{6}}p_{_{-}}^2 p_{z}/2(m+E)\cr\cr\cr 
(4 p_{_{-}} p_{z}^2 + p_{_{-}}^2
p_{_{+}} + 2m(m+E)p_{_{-}})/2(m+E)\cr\cr\cr 
-(2 p_{z}^3 + (p_{_{-}} p_{_{+}}
+2m(m+E))p_{z})/(m+E)\cr}.
\eqn\jbpgfe$$ 
}
\endpage
The antiparticle covariant spinors are
$$v_\sigma(\v p\,) = \gamma_5~ u_\sigma(\v p\,).
\eqn\jbpgg$$
Just as in the previous cases, for a massless particle travelling 
along the $\hat z$ direction,
only  $u_{\pm{ 2}}(p_z)$ and $v_{\pm{2}}(p_z)$ are found to be non-null.
The $u_{\pm{1},0}(p_z)$ and $v_{\pm{1},0}(p_z)$ vanish identically. While
$u_{+{2}}(E)$ and $v_{+{2}}(E)$ are identical, the $u_{-{2}}(p_z)$ $v_{-{2}}
(p_z)$
again differ by a sign.

\section{Orthonormality of $(j,0)\oplus(0,j)$ Covariant Spinors}

We define $\gamma_{\circ}^{\smalltype CA}$ as the obvious generalization
of the Dirac $\gamma_{\circ}$
$$\gamma_{\circ}^{\smalltype CA}=\pmatrix{I&0\cr 0&-I},
\eqn\jbpfa$$
and introduce (in the canonical representation)
$$\overline \psi_\sigma(\v p\,)= \psi
^\dagger_\sigma(\v p\,)  \gamma_{\circ}^{\smalltype CA}.
\eqn\jbpfb$$ 
To ensure the correctness of the $j=1,\,3/2,$ and $j=2$ 
particle/antiparticle covariant
$u_\sigma(\v p\,)$ and $v_\sigma(\v p\,)$ presented here, 
we have through brute force
matrix multiplication verified that
$$\overline u_\sigma(\v p\,)~u_{\sigma'}(\v p\,) = m^{2j}\delta_{\sigma\sigma'}
\eqn\jbpfca$$
$$\overline v_\sigma(\v p\,)~v_{\sigma'}(\v p\,) = -m^{2j}\delta_{\sigma\sigma'}
\eqn\jbpfcb$$
In the canonical representation the origin of the ``minus'' sign in the $rhs$ of
the orthonormality condition \jbpfcb ~can be readily traced back to the
structure of 
$\gamma_{\circ}^{\smalltype CA}$, 
and the fact that $v_\sigma(\v p\,)$ are obtained 
from the $u_\sigma(\v p\,)$ via the
matrix $\gamma_5$. Symbolically, we have
$$u\sim\pmatrix{a\cr b},~\overline u\sim \pmatrix{a^\ast& b^\ast}
\pmatrix{I&0\cr 0&-I}=
\pmatrix{a^\ast&- b^\ast}
\eqn\jbpfd$$
Hence
$$\overline u~u\sim a^\ast a-b^\ast b.\eqn\jbpfe$$
While for $v$
$$v\sim \gamma_5~u \Rightarrow v=\pmatrix{b&a} \Rightarrow \overline v~v\sim
b^\ast b-a^\ast a = -\overline u~u, ~~~~QED.
\eqn\jbpff$$
The (relative) minus sign in the $rhs$ of the orthonormality relations \jbpfca~
and \jbpfcb~ is essential for the existence of a conserved charge constructed
from the field operators made out of these spinors.

\chapter{Causal Propagators for $(j,0)\oplus(0,j)$ Fields}

From the covariant spinors, we can construct field operators. The same
arguments [\Thesis] which apply for the Dirac case hold here. The field 
operator for the $(j,0)\oplus(0,j)$ matter fields is
$$\eqalign{ \Psi^{(j,0)\oplus(0,j)}&(x) = 
\sum_{\sigma=-j}^{+j} \int {d^3p\over (2\pi)^{3} } {1\over 2\,\omega_{\v p}}
 \cr
& \times\Big[ u_\sigma(\vec p\,)\, a(\vec p, \sigma)\, \exp(-i p\cdot x)
+  v_\sigma(\vec p\,) \,b^\dagger(\vec p, \sigma) \,\exp(+i p\cdot x) \Bigr ]
,\cr}\eqn\fieldoperator$$
with
$$\omega_{\v p\,} = \sqrt{m^2 + {\v p\,}^2}, ~~
{\overline
\Psi}^{(j,0)\oplus(0,j)}(x)\equiv{\Psi^{(j,0)\oplus(0,j)}}^\dagger(x)\,
\gamma^{\smalltype CA}_{\circ}\,\,\,.
\eqn\psibar$$ 

The object which enters a perturbation calculation as the propagator
is the vacuum expectation value of the time--ordered field operators,
$$\la x\vt\,{\cal S}^{j}_{ FD}\,\vt y \ra \equiv
\la~~\vt\, T[\Psi^{(j,0)\oplus(0,j)}(x)~\overline\Psi^
{(j,0)\oplus(0,j)}(y)]\,\vt~~\ra\,\,\,. \eqn\time$$ 
This propagator is not equal to the Green's function for the Joos--Weinberg
equations. This is because these equations support [\PRCb--\Vanderbilt] spurious
and unphysical solutions. A Green's function constructed from these equations
would propagate these extra solutions while \time~will propagate only
the physical solutions.
Using $\{a_\sigma(\v p\,)\,,\,a^\dagger_{\sigma '}({\v p\,}'
\,)\}=(2\pi)^3\, 2\omega_{\v p}\,
\delta_{\sigma\sigma'}\delta(\v p-{\v p\,}')$, for fermions and the
similar relation
for bosons (with the anticommutator replaced by a commutator), we obtain the
configuration space Feynman--Dyson propagator for arbitrary spin,
$$\eqalign{\la x & \vt\,{\cal S}^{j}_{ FD}\,\vt y \ra
\,=\,\sum_{\sigma=-j}^{+j}\int {d^3p\over(2\pi)^3} {1\over 2\omega_{\v p}}\cr 
&\times\Big[u_\sigma(\v p\,)\,{\overline u}_\sigma(\v p\,) \,e^{-ip\cdot(x-y)}
\,\theta(x^\circ-y^\circ) + \epsilon\,
v_\sigma(\v p\,)\,{\overline v}_\sigma(\v p\,)\, e^{+ip\cdot(x-y)}
\,\theta(y^\circ-x^\circ)\Big],\cr}\eqn\configuration$$
with
$$\epsilon\,=\,\cases{+1& for bosons,\cr
                  -1& for fermions.\cr}\eqn\ep$$
The momentum--space Feynman--Dyson propagator is given by
$$\eqalign{ \langle k'\vert\,S_{FD}^{j}\,\vert k\rangle&=\int {d^4x\over (2\pi
)^3} 
{d^4y\over (2\pi )^3}\,e^{ik'\cdot x}\,e^{-ik\cdot y}\langle x \vert
\,{\cal S}^j_{FD}\,\vert y\rangle\cr
&= -{i\,\delta^{(4)}(k'-k)\over{(2\pi)^2\,2\omega_{\v k}}}
\sum_{\sigma = -j}^{+j}
{\Bigg(} 
{ u_\sigma (\vec k\,)
\,\overline{u}_\sigma (\vec k\,) \over k_o+i\eta -E(\vec k\,)}
-\epsilon\,{v_\sigma 
(-\vec k\,)\,\overline{v}_\sigma (-\vec k\,)\over k_o-i\eta
+E(\vec k\,)   } {\Bigg)}. \cr}\eqn\momentum$$
The propagator has the structure of a typical particle--hole propagator
in non--relativistic quantum mechanics. Although not a common expression,
the standard Feynman propagator for spin 1/2 
[\Charly] can also be written in the form \momentum.

Recently [\Waveeqn] we have inverted the propagator given by Eq. \momentum~ and
established that the resulting wave equation propagates only the kinematically
acceptable solutions. Weinberg [\WeinbergA], on the other hand, added certain
contact terms to the $r.h.s.$ of \time. The resulting Weinberg  equation,
though {\it manifestly} covariant, propagates kinematically spurious solutions
as we have shown in a recent publication [\brief].

The only other element
needed in perturbative calculations is an appropriate model interaction.
We will treat model interactions in detail as we make applications.
Phenomenological interactions involving an even number of 
$(j,0)\oplus(0,j)$ matter fields and a scalar,  pseudoscalar
or  vector field are straightforward. We here give some simple
examples.

First, the coupling of a scalar with two particles of spin $j$ can
be written as
$$\eqalign{ {\cal L}(x)\,=\,g_1
\,\Phi_s(x)\,&\overline\Psi^{(j,0)\oplus(0,j)}(x)\, 
\Psi^{(j,0)\oplus(0,j)}(x) \cr
&\, +\,g_2\, \pl^\mu\Phi_s(x)\, \overline\Psi^{(j,0)\oplus(0,j)}(x)\,
\pl_\mu\Psi^{(j,0)\oplus(0,j)}(x), \cr}\eqn\Laga$$
where $\Phi_s(x)$ is the field 
operator associated with the scalar particle and $g_1$ and $g_2$ are
coupling constants to be determined experimentally.
Secondly, the coupling of a pseudoscalar with two particles of spin $j$ 
could be written as
$$\eqalign{
{\cal L}(x)\,=\,\eta_1\, \Phi_p(x)\,&\overline\Psi^{(j,0)\oplus(0,j)}(x)\,
\gamma^5\,
\Psi^{(j,0)\oplus(0,j)}(x)\cr
& \,+\,
\eta_2\, \pl^\mu\Phi_p(x)\,\overline\Psi^{(j,0)\oplus(0,j)}(x)\,\gamma^5\,
\pl_\mu\Psi^{(j,0)\oplus(0,j)}(x),\cr} \eqn\Lagb$$
where $\Phi_p(x)$ is the field 
operator associated with the pseudoscalar particle and 
$\eta_1$ and $\eta_2$ are 
coupling constants.

Finally, the interactions associated with two--photon production of a spin-2
meson such as the $f_2(1720)$ 
may be postulated to be of the form
$$\eqalign{
{\cal L}(x)\,=\,\alpha_\circ\, A^\mu(x)\,& \overline\Psi^{(2,0)\oplus(0,2)}(x)\,
\pl_\mu\Psi^{(2,0)\oplus(0,2)}(x)\cr
&\, +\,\sum_{\{P\}}
\alpha_{\{P\}} \,
A^\mu(x)\, \overline\Psi^{(2,0)\oplus(0,2)}(x)\,{\gamma_{\mu\nu}}^{\nu\lambda}\,
\pl_\lambda\Psi^{(2,0)\oplus(0,2)}(x),\cr}\eqn\lagc$$
where the summation on $\{P\}$ is a sum on the permutations of the order of the
indices of ${\gamma_{\mu\nu}}^{\nu\lambda}$.
The  $\gamma_{\mu\nu\lambda\sigma}$ are a set of $10\times 10$ matrices,
 related to the spin--2 gamma matrices of Weinberg [\WeinbergA] by
$$\gamma_{\mu\nu\sigma\lambda}\,=\,A\,
\gamma_{\mu\nu\sigma\lambda}^{Weinberg}\,A^{-1}\,\eqn\gammad$$ 
with $A$ given in Eq. \A. The explicit form of
$\gamma_{\mu\nu\sigma\lambda}$ can be found in [\PRCb] and [\Thesis]. 
The construction of interactions which involve an
odd number of $(j,0)\oplus(0,j)$ matter fields of a given $j$
is a little more involved [\WeinbergZ]. 

\chapter{Conclusions}

A totally satisfactory quantum field theory of particles with high spin
does not yet exist. This makes it difficult to treat high--spin particles
in a covariant manner. We have here taken a different attitude to this
subject than seems to have been previously adopted. Not having been able to
resolve the difficulties with the existing theories, and indeed having found
new problems [\PRCb--\Vanderbilt], we address the problem from a pragmatic point
of view --- can one build a covariant phenomenology of high--spin particles
which is internally consistent? We here make the proposal of doing this
by avoiding any explicit reference to a wave equation and constructing
the individual elements needed for a perturbation theory. The first of
these, the covariant spinors, can be constructed following the work
of Wigner [\Wigner] and Weinberg [\WeinbergA]. We have here reviewed
their work and provided a practical and detailed technique (a generalization
of the spin one--half approach of Ryder [\Ryder]) for generating the
spinors. Explicit expressions for spinors with $j=~$1, 3/2 and 2 are
given. Although the algebra becomes increasingly tedious, the approach can 
be continued to higher $j$. Field operators can be constructed from
the spinors in complete analogy to the spin one--half case. The
free--particle propagator is then defined in terms of the vacuum
expectation value of the time ordered field operators. We provide
explicit expressions for these propagators. The propagator which we
define here is not equal (except for the $j=~$1/2 case) to the Green's
function of the Joos--Weinberg equations. Finally, we have provided some
simple examples of model interactions. This provides all of the
necessary ingredients to formulate a perturbation theory and thus
can form the basis for a phenomenological approach to the interactions
of high--spin particles. 

Several additional points need to be mentioned. First, although we
propose to define our theory as equal to the perturbation theory,
calculations need not be done order by order in the expansion. For example,
classes of diagrams can be summed through infinite order by making use
of integral equations and solving them numerically. The actual implementation
of the approach proposed here must be tailored to the particular physical 
system under investigation. Secondly, if we include phenomenological form
factors, all matrix elements will be finite. This does not, however,
remove the necessity of renormalizing masses and coupling constants.
The calculations need to be executed in terms of physical masses
and measurable coupling constants. This approach has been successful
for building a phenomenology of the pion--nucleus interaction [\Meson].
We expect that working without a wave equation and an underlying Lagrangian
will, at some point, limit the scope of problems which we can undertake.
In the mean time, we are examining several applications and have yet
to encounter any basic limitations.

\endpage
\Appendix{A} 
\noindent
{\underbar{Three Rotations}} \rm about each of the $(x,y,z)$--axes. The
transformation matrices relating $x^{\prime\mu}$ with $x^\mu$,
$x^{\prime\mu}={R^\mu}_\nu x^\nu$, 
are given by
$$[{R^\mu}_\nu(\theta_x)]=\pmatrix{1&0&0&0\cr
                                   0&1&0&0\cr
                                   0&0&\cos(\theta_x)&-\sin(\theta_x)\cr
                                   0&0&\sin(\theta_x)&\cos(\theta_x)\cr},
\eqn\apd$$

$$[{R^\mu}_\nu(\theta_y)]=\pmatrix{1&0&0&0\cr
                              0&\cos(\theta_y)&0&\sin(\theta_y)\cr
                              0&0&1&0\cr
                              0&-\sin(\theta_y)&0&\cos(\theta_y)\cr},
\eqn\ape$$

$$[{R^\mu}_\nu(\theta_z)]=\pmatrix{1&0&0&0\cr
                              0&\cos(\theta_z)&-\sin(\theta_z)&0\cr
                              0&\sin(\theta_z)&\cos(\theta_z)&0\cr
                              0&0&0&1\cr}.
 \eqn\apf$$
\noindent
$[{R^\mu}_\nu(\theta_i)]$ represents a rotation by
$\theta_i$ about the $ith$--axis. The rows and columns are
labelled in the order $0,1,2,3$.\goodbreak

\noindent
{\underbar{Three Lorentz Boosts}} along each of the $(x,y,z)$--axes. 
The boost matrix 
for a boost along the positive direction of the unprimed $x$-axis, by velocity
\footnote{4}{This is the velocity which a particle at rest in the unprimed
frame acquires when seen from the primed frame.} 
$v$, is given by 
$$[{B^\mu}_\nu(\varphi_x)]=\pmatrix{\cosh(\varphi_x)&\sinh(\varphi_x)&0&0\cr
                   \sinh(\varphi_x)&\cosh(\varphi_x)&0&0\cr
                   0&0&1&0\cr
                   0&0&0&1\cr},
\eqn\apg$$
\noindent
with $x^{\prime\mu}={B^\mu}_\nu(\varphi_x) x^\nu$. Similarly

$$[{B^\mu}_\nu(\varphi_y)]=\pmatrix{\cosh(\varphi_y)&0&\sinh(\varphi_y)&0\cr
                     0&1&0&0\cr
                     \sinh(\varphi_y)&0&\cosh(\varphi_y)&0\cr
                     0&0&0&1\cr},
\eqn\aph$$

$$[{B^\mu}_\nu(\varphi_z)]=\pmatrix{\cosh(\varphi_z)&0&0&\sinh(\varphi_z)\cr
                     0&1&0&0\cr
                     0&0&1&0\cr
                     \sinh(\varphi_z)&0&0&\cosh(\varphi_z)\cr}.
\eqn\api$$

\Appendix{B}
\noindent

Here we provide expansions for $\cosh(2\v J\cdot\v\varphi)$
and $\sinh(2\v J\cdot\v\varphi)$.
In the identities below we have defined $\eta=(2\v J\cdot\hat p\,)$\goodbreak
\noindent
{\smalltype INTEGER SPIN:}\rm\goodbreak
$$\cosh(2\v J\cdot\v\varphi)=1+\sum_{n=0}^{j-1}
{{(\eta^2)(\eta^2-2^2)(\eta^2-4^2)\ldots(\eta^2-(2n)^2)}
\over
{(2n+2)!}}\sinh^{2n+2}\varphi,\eqn\zbpa$$

$$\sinh(2\v J\cdot\v\varphi)=\eta\cosh\varphi~\sum_{n=0}^{j-1}
{{(\eta^2-2^2)(\eta^2-4^2)\ldots(\eta^2-(2n)^2)}
\over
{(2n+1)!}}\sinh^{2n+1}\varphi.\eqn\zbpb$$
\endpage
\noindent
{\smalltype HALF INTEGER SPIN:}\rm\goodbreak
$$\cosh(2\v J\cdot\v\varphi)=\cosh\varphi~\Biggl[1+\sum_{n=1}^{j-1/2}
{{(\eta^2-1^2)(\eta^2-3^2)\ldots(\eta^2-(2n-1)^2)}
\over{(2n)!}}\sinh^{2n}\varphi\Biggr],\eqn\zbpc$$
$$\sinh(2\v J\cdot\v \varphi)=\eta\sinh\varphi~\Biggl[1+\sum_{n=1}^{j-1/2}
{{(\eta^2-1^2)(\eta^2-3^2)\ldots(\eta^2-(2n-1)^2)}
\over{(2n+1)!}}\sinh^{2n}\varphi\Biggr].\eqn\zbpd$$

\endpage

\centerline{Table I}
\vskip 1.0in
\vbox{
\hrule\hrule
\medskip
\settabs 6 \columns
\+ { Rotation about:}& & & { Boost along:} & & \cr
\vskip 0.1in
\+ { x--axis} & { y--axis} & { z--axis} &{ x--axis} & 
{ y--axis} & { z--axis}\cr
\medskip
\hrule
\medskip
\+ $\lambda^{23}=-\lambda^{32} $ & 
$\lambda^{31}=-\lambda^{13} $ &
$\lambda^{12}=-\lambda^{21} $  & 
$\lambda^{10}=-\lambda^{01} $ &
$\lambda^{20}=-\lambda^{02} $ & 
$\lambda^{30}=-\lambda^{03} $\cr
\+ $=\theta_x$ & 
$=\theta_y$ &
$=\theta_z$  & 
$=\varphi_x$ &
$=\varphi_y$ & 
$=\varphi_z$\cr
\medskip
\hrule\hrule
}
\vskip .5truein

\noindent
Table I. {\it Nonvanishing} $\lambda^{\mu\nu} = {\lambda^\mu}_\epsilon
\eta^{\epsilon\nu}$. Not we only tabulate the nonvanishing $\lambda^{\mu\nu}$,
as such, for example $\lambda^{\mu\ne 2\,\nu\ne 3}=
-\lambda^{\nu\ne 3\,\mu\ne 2= 0}$ for a rotation about x-axis. Similar
comments apply for other transformations.

\endpage\par\penalty-400\vskip\chapterskip\spacecheck\referenceminspace
   \ifreferenceopen \Closeout\referencewrite \referenceopenfalse \fi
   \line{\fourteenrm\hfil REFERENCES\hfil}\vskip\headskip
   \input referenc.txa
   \bye